\documentclass[fleqn,usenatbib]{mnras}

\usepackage{widetext}
\usepackage{graphicx}
\usepackage{dcolumn}
\usepackage{bm}
\usepackage{natbib}
\usepackage{multirow}
\usepackage{url}
\usepackage{color}
\usepackage{xcolor}
\usepackage{tikz}
\usepackage{float}
\usepackage{bbold}
\usepackage{cancel}
\usepackage{amssymb}
\usepackage{amsmath}

\usepackage{colortbl}

\usepackage{breakurl}
\usepackage{comment}

\newcommand{\mbi}[1]{\mbox{\boldmath$#1$}}

\newcommand{\mat}[1]{\mbox{\rm #1}}
\newcommand{\lsim}{\mbox{${\,\hbox{\hbox{$ < $}\kern -0.8em \lower 1.0ex\hbox{$\sim$}}\,}$}}
\newcommand{\gsim}{\mbox{${\,\hbox{\hbox{$ > $}\kern -0.8em \lower 1.0ex\hbox{$\sim$}}\,}$}}

\newcommand{\dd}{{\rm d}}

\def\beqn{\vspace{2mm}
\begin{eqnarray}} 
\def\eeqn{\vspace{2mm} 
\end{eqnarray}}

 \def\la{\langle}

\newcommand{\be}{\begin{equation}}
\newcommand{\ee}{\end{equation}}
\newcommand{\ba}{\begin{eqnarray}}
\newcommand{\ea}{\end{eqnarray}}
\newcommand{\brr}{\begin{array}}
 
\newcommand{\err}{\end{array}}
\newcommand{\bc}{\begin{center}}
\newcommand{\ec}{\end{center}}

\newcommand{\hinvm}{\,h^{-1}\,{\rm Mpc}}
\newcommand{\hminv}{\,h\,{\rm Mpc}^{-1}}

\title[Cosmic web and halo distribution]{The cosmic web connection to the dark matter  halo distribution through gravity} 

\author[F.-S.~Kitaura et al.]{F.-S.~Kitaura$^{1,2}$\thanks{E-mail: \href{mailto:fkitaura@iac.es}{fkitaura@iac.es}}, A. Balaguera-Antol\'{\i}nez$^{1,2}$, F. Sinigaglia$^{1,2,3}$ \& M. Pellejero-Ib\'a{\~n}ez$^{4}$ \\ \\ $^1$Instituto de Astrof\'{\i}sica de Canarias (IAC), Calle V\'{\i}a Lactea s/n, 38200, La Laguna, Tenerife, Spain \\  $^2$Departamento de Astrof\'{\i}sica, Universidad de La Laguna (ULL), E-38206, La Laguna, Tenerife, Spain\\  $^{3}$Department of Physics and Astronomy, Università degli Studi di Padova, Vicolo dell’Osservatorio 3, I-35122, Padova, Italy\\ $^{4}$Donostia International Physics Centre, Paseo Manuel de Lardizabal 4, 20018 Donostia-San Sebastian, Spain}

\date{Accepted XXX. Received YYY; in original form ZZZ}
\pubyear{2019}

\begin{document}
\label{firstpage}
\pagerange{\pageref{firstpage}--\pageref{lastpage}}
\maketitle

\begin{abstract}

This work investigates the connection between the cosmic web and the halo distribution through the  gravitational potential at the field level. 
We combine three fields of research, cosmic web classification, perturbation theory expansions of the halo bias, and halo (galaxy) mock catalogue making methods. In particular, we use the invariants of the tidal field  and the velocity shear tensor as generating functions to reproduce the halo number counts of a reference catalogue from full gravity calculations, populating the dark matter field on a mesh well into the non-linear regime ($3\hinvm$ scales). Our results show an unprecedented agreement with the reference power spectrum within  1\% up to  $k=0.72\hminv$. By analysing the three-point statistics on large scales (configurations of up to  $k=0.2\hminv$), we find  evidence for non-local bias at the 4.8 $\sigma$ confidence level, being  compatible with the reference catalogue. In particular, we find that a detailed description of tidal anisotropic clustering on large scales is crucial to achieve this accuracy at the field level.
These findings can be particularly important for the analysis of the next generation of galaxy surveys  in mock galaxy production.

\end{abstract}


\setcounter{footnote}{0}

\section{Introduction}

\noindent
  The cosmic web  represents the  complex large scale patterns observed in the late time Universe \citep[e.g.][]{2009LNP...665..291V}. These are dominated by a filamentary network, connected through knots, which leave large void regions delimited by sheet-like structures. 
  Applications of the cosmic web range from using filaments to study the missing baryons in the inter-galactic medium  \cite[][]{2015ApJ...806..113G,2015Natur.528..105E,2019A&A...624A..48D}, over using voids to study baryon acoustic oscillations (BAOs)  \citep[][]{2016PhRvL.116q1301K,2016MNRAS.459.4020L,2020MNRAS.491.4554Z}, to study redshift space distortions (RSD) \citep[][]{2017PhRvD..95f3528C,nadathur18}, to constrain dark energy \citep[][]{2007PhRvL..98h1301P,2010MNRAS.403.1392L,2012MNRAS.426..440B,2015PhRvD..92h3531P},  or to challenge gravity \citep[][]{2019MNRAS.484.5267K}.
  Moreover, the formation of the cosmic web is intricately related to galaxy formation and evolution, which can be studied through the location and orientation of galaxies and their properties in the cosmic web \citep[see e.g.][]{2001ApJ...557..117B,2004ApJ...615L.101B,2004MNRAS.353..713K,2005ApJ...629..143B,2006MNRAS.373..469B,2007ApJ...655L...5A,2008ApJ...688...78L,2008MNRAS.383..907B,2012MNRAS.421L.137L,2013AJ....145..120B,2014MNRAS.445..988N,2015ApJ...802...82F,2017ApJ...848...60Y,2018MNRAS.473.2486S,2018MNRAS.473.3941F,2019MNRAS.485.2367C,2019MNRAS.483.4501A,2019OJAp....2E...7A,2019MNRAS.482.1900H,2019MNRAS.487.1607G}.
  Each cosmic web type has its own theoretical framework and dedicated observing and analysis strategies 
 \citep[][]{2004MNRAS.350..517S,2008LNP...740..335V,2011MNRAS.414..384S,2012MNRAS.421..926P,2014MNRAS.438.3465T,2016IAUS..308..591K}.
   Recent studies aim at exploiting the cosmic web to constrain cosmological parameters \citep[][]{2018arXiv180804728M,Fang_2019,Naidoo_2019}.
     For all these reasons,  understanding and characterising the cosmic web represents one of the major efforts in modern cosmology.

\noindent 
However, the classification of the cosmic web is still arbitrary, and a long standing debate discusses which method is optimal to determine the different types of structures in the matter density field. A comprehensive comparison of the different methods is presented in \citet[][]{2018MNRAS.473.1195L}.
While there is no doubt about the scientific interest to study the cosmic web, and its usefulness has already been proven, a more quantitative criterion relying on fundamental principles is still missing.
The cosmic web is usually directly defined based on the dark matter field. But another approach could consist of defining it based on its tracer, which represents the actual observable.
To this end, we focus in this work on the halos hosting the galaxies, which span the cosmic web.
In fact, the theoretical pillars for the current understanding of the formation of halos have been laid long ago, and spectacular progress has been made in the past decades \citep[see][and references therein]{2018PhR...733....1D}.
Their connection to the density field has been thoroughly modelled, and general functional non-linear and non-local dependencies have been studied  (see Sec.~\ref{sec:theory}).
We suggest therefore to take the perspective of dark matter halos, to revisit the definition of the cosmic web. 
The accuracy at which we are able to reproduce the summary statistics of the distribution of halos, depending on the specification of the elements, which presumably determine the cosmic web, provides us a measure  of the quality of the cosmic web definition itself. This procedure permits us to reduce the number of relevant quantities derived from the gravitational potential to only a few.

\noindent 
A series of papers inspire this work, ranging from the  cosmic web definition  \citep[][]{1970A&A.....5...84Z,1996Natur.380..603B,2007MNRAS.375..489H}, over the study of halo bias \citep[][]{1986ApJ...304...15B,2002MNRAS.329...61S,2009JCAP...08..020M,2012PhRvD..85h3509C}, 
to the generation of halo catalogues.
A precise understanding of halo bias can be key to produce fast and  accurate mocks, which then can be used to study systematics in galaxy catalogues, and to put errors bars on the cosmological information  (for a variety of mock making methods see \citep[][]{1996ApJS..103....1B,2002MNRAS.329..629S,2002MNRAS.331..587M,2014MNRAS.439L..21K,2014MNRAS.437.2594W,2015MNRAS.446.2621C,2015MNRAS.450.1856A,2016MNRAS.459.2118K,2016MNRAS.459.2327I,2017MNRAS.472.4144V,2019MNRAS.483.2236S}, and for a comparison between them  see \citep[][]{2015MNRAS.446.2621C,2019MNRAS.485.2806B,2019MNRAS.482.1786L,2019MNRAS.482.4883C,2020MNRAS.493..586P}). 
In the present work we combine the different fields of research to test in detail, how the cosmic web information can be used within the framework of perturbation theory (PT) halo bias to reproduce the clustering statistics of full gravity calculations.

In the remainder of this paper, we  first recap the theoretical background to the classical cosmic web definition, and the general perturbative bias expansion. Subsequently,  we  present the connection between both, and motivated by that, we  propose a new cosmic web classification based on the invariants of the tidal field or velocity shear tensor. Then we briefly describe the reference $N$-body simulations and our machine learning approach \texttt{BAM}. In the section after, we  present numerical studies based on $N$-body simulations to quantify the information gain with our proposed cosmic web classification. Finally, we  present our conclusions.

\section{Theoretical background}
\label{sec:theory}

In this section, we  revise first the definition of the cosmic web, subsequently the perturbation halo bias, and finally we  present a unified picture connecting both.

\subsection{The cosmic web picture}
\label{sec:cw}

\noindent 
The cosmic web arises from the growth of structures starting from approximately Gaussian density perturbations \citep[see e.g.][]{2010gfe..book.....M}. The connection between the primordial and the final density fluctuations can be described through the mapping between Lagrangian to Eulerian space, which central role is played by the tidal field tensor  $\mathcal T_{ij}\equiv\partial_i\partial_j\phi$ constructed from the gravitational potential $\phi$  with $i$ and $j$ standing for different spatial directions \citep[see the pioneering work by][]{1970A&A.....5...84Z}.

Let us start by considering the mapping between 
the initial (at initial redshift $z_0$) and the final (at final $z$) coordinates 
of test particles. In Lagrangian perturbation theory this relation is expressed via a displacement
field, ${\vec \psi}({\vec q})$ \citep[see][for a review]{2002PhR...367....1B}:
\be
\label{eq:lag}
 \vec r = \vec{q} + \vec{\psi}(\vec{q}) 
 \, ,
\ee
which defines a unique mapping between $\vec q$
and $\vec r$ (usually referred to as Lagrangian and Eulerian coordinates).
In addition, one needs to consider the Poisson equation relating the gravitational potential $\phi$ to the over-density field $\delta$ through the Laplacian operator:
\be
\nabla^2\phi(\vec q)=\delta(\vec q) 
\,,
\ee
where $\delta\equiv\rho/\bar{\rho}-1$, and $\rho$ is the (dark) matter density field.
If we further assume that the test particles were initially
homogeneously distributed, then we can write the following mass conservation relation: 
\be
\rho(\vec r)\dd \vec r=\langle \rho(z_0)\rangle \dd \vec{q}
\,.
\ee
The inverse of the Jacobian of the coordinate transformation defines the
over-density field:
\be
\label{eq:jac}
1+\delta(\vec r(\vec q))=\mat J(\vec q)^{-1}  \,,
\ee
with
\be
\mat J(\vec q) \equiv \left|\frac{\partial\vec r}{\partial\vec q}\right| \,.
\ee
The displacement field is given by the gradient of a potential (since any initial curl decays with cosmic evolution considering up to second order Lagrangian perturbation theory) $\vec\psi(\vec q)=-\nabla \phi(\vec q)$, and hence Eq.~(\ref{eq:jac}) can be rewritten as:
\be
\label{eq:deltapsi}
1+\delta(\vec q)={\left|\delta^{\rm K}_{ij}+\partial_{i}\partial_{j}\phi(\vec q)\right|^{-1}}\,.
\ee
We can define the tidal field, strain, or  deformation
tensor
\be
\label{eq:tidal}
\mathcal T_{ij}(\vec q)\equiv\partial_{i}\partial_{j}\phi(\vec q)\,,
\ee
which
is symmetric, allowing us to diagonalize it
with eigenvalues $\lambda_1\geq\lambda_2\geq\lambda_3$.
This permits us to rewrite  Eq.~(\ref{eq:deltapsi}) as:
\be
\label{eq:deltalambda}
1+\delta(\vec q)={\left[\left(1-\lambda_1\left(\vec q\right)\right)\left(1-\lambda_2\left(\vec q\right)\right)\left(1-\lambda_3\left(\vec q\right)\right)\right]^{-1}}\,,
\ee
Linearly expanding Eq.~(\ref{eq:deltalambda}) we obtain the linear density field in Lagrangian coordinates:
$\delta(\vec q)\equiv\lambda_1(\vec q)+\lambda_2(\vec q)+\lambda_3(\vec q)$. 
By linearising Eq.~(\ref{eq:deltapsi}) we obtain the Zel'dovich approximation: 
\be
\label{eq:zeld}
\delta(\vec q)\simeq-\nabla\cdot\vec\psi(\vec q)\,.
\ee
The Zel'dovich approximation is often identified with Eq.~(\ref{eq:deltalambda}), instead of linear Lagrangian perturbation theory, as  we do here with Eq.~(\ref{eq:zeld}). However, Eq.~(\ref{eq:deltalambda}) is more general and led to a structure formation model  (which we discuss below) and admits higher order Lagrangian perturbation theory expressions  
\citep[][]{1992ApJ...394L...5B,1993MNRAS.264..375B,1994MNRAS.267..811B,1995A&A...296..575B,1995MNRAS.276..115C,2000ApJ...534L.117B}.
In the approximation of curl-free velocity fields, they can be directly inferred from the normalised divergence of the density field 
 \be
 \theta \left(\vec r\right)\equiv-\left(a~H(a)~d\ln D(a)/d\ln a\right)^{-1}\nabla\cdot\vec  v\left(\vec r\right)\,.
 \ee
 In the linear velocity  approximation $\theta=\delta$.
From an Eulerian perspective, one needs to consider the inverse mapping $\vec q=\vec r-\vec \psi(\vec q(\vec r))$ \citep[][]{2012MNRAS.425.2443K}, which under mass conservation  yields  the inverse Jacobian: 
$\mat J'(\vec r) \equiv \left|\frac{\partial\vec q}{\partial\vec r}\right|$, 
and hence to 
$1+\delta(\vec r)=\left|\delta^{\rm K}_{ij}-\partial_{i}\partial_{j}\phi(\vec r)\right|$. 
Once shell crossing allows for multi-streaming \citep[e.g.][]{2015MNRAS.454.3920H}, there is no unique solution without the peculiar velocity information, unless some  approximation is done to ensure reversibility \citep[e.g.][]{1992ApJ...391..443N,1993ApJ...405..449G,2012MNRAS.425.2443K,2018PhRvD..97b3505S}, some optimisation is applied \citep[e.g.][]{1989ApJ...344L..53P,2000MNRAS.313..587N,2002Natur.417..260F}, or an   ensemble of solutions in a statistical sense is sought \citep[e.g.][]{2013MNRAS.429L..84K,2013MNRAS.432..894J,2013ApJ...772...63W}. 

In the linear regime one obtains  the same approximations, as given by Eqs.~(\ref{eq:deltalambda}) and (\ref{eq:zeld}), substituting the $\vec q$-coordinates with the $\vec r$-coordinates dependence.
For what matters, the same theoretical grounds are valid, and from now on, we  consider the tidal field tensor expressed in Eulerian coordinates.

In particular, the density field in Eulerian coordinates (computed from the non-linear gravitational potential) reads:
\be
\delta(\vec r)\equiv\lambda_1(\vec r)+\lambda_2(\vec r)+\lambda_3(\vec r)\,.
\ee

The result shown in  Eq.~(\ref{eq:deltalambda})
led to the top-down  scenario of structure formation, in which the catastrophies of infinite density occurred first along the largest eigenvalue to first form sheets also known as Zel'dovich "pancakes", then along the second largest to form filaments, and finally along the smallest eigenvalue to form knots   \citep[][]{2014MNRAS.437.3442H,2018JCAP...05..027F}.

This  is in contrast with the well  established hierarchical model scenario, in which smaller objects form first and by merging processes grow to larger ones  \citep[][]{1974ApJ...187..425P,1978ApJ...221...19F,1978MNRAS.183..341W,1986ApJ...304...15B}.
In fact, Eq.~(\ref{eq:deltalambda}) can be used to compute gravitational collapse with LPT and combined with the hierarchical model to generate halo distributions including accurate merger histories \citep[][]{1996ApJS..103....1B,2002ApJ...564....8M}. \citet{1996Natur.380..603B} 
investigated both scenarios studying the appearance of sheets (``pancakes") as compared to filaments in both $N$-body and  Zel'dovich calculations, finding a preference for the latter. 
That paper coined the term cosmic-web, and used arguments based on eigenvalues of the shear of the velocity field
\be
\Sigma_{ij}\equiv \frac{1}{2}\left( \partial_iv_j+\partial_jv_i\right) 
\ee

(which is equivalent to the tidal field in linear theory) to make a cosmic web classification.  
They considered the  ellipticity    
$e\propto(\lambda_1-\lambda_3)$ and the  prolatness $p\propto(\lambda_1+\lambda_3-2\lambda_2)$ (both up to a consistent normalisation of typically $2\delta$) to define filaments: $p\sim-e$ and sheets: $p\sim e$, which imply $\lambda_1\sim\lambda_2$ and  $\lambda_2\sim\lambda_3$, respectively.
A more systematic  cosmic web classification, also inspired by the Zeldovich ``pancakes" formation, was introduced by \citet{2007MNRAS.375..489H}, commonly known as the T-web. In this classification, they considered knots, as the regions in which gravitational collapse causes matter inflow  expressed  through positive eigenvalues (knots: $\lambda_1,\lambda_2,\lambda_3>0$); voids as opposed to knots are described as  expanding regions with negative eigenvalues (voids: $\lambda_1,\lambda_2,\lambda_3<0$); filaments being closer to knots are defined as regions with two positive eigenvalues, and a negative one (filaments: $\lambda_1,\lambda_2>0; \lambda_3<0$); while sheets are defined by one  positive, and two negative eigenvalues (sheets: $\lambda_1>0;\lambda_2, \lambda_3<0$). The cases in which any  eigenvalue coincides with zero has been neglected, as it can be arbitrarily  assigned to any limiting case.  We note, that this picture coincides with the one provided by \citep[][]{1996Natur.380..603B}, since also in the T-web definition filaments have the largest eigenvalues closer to each other, while for sheets this happens for the lowest ones.  

These definitions, however  are more qualitative than quantitative, since there is not a first principle guaranteeing that a filament is really such a cosmic web type. In fact, this cosmic web classification depends on the smoothing scale, mesh resolution, and mass assignment scheme. To alleviate this, an eigenvalue threshold (instead of zero) was introduced \citep[][]{2009MNRAS.396.1815F}, and a multi-scale classification was developed  \citep[][]{2007A&A...474..315A,2013MNRAS.429.1286C}.
Alternatively, the velocity shear tensor  has been revived to classify the cosmic web \citep[V-web:][]{2012MNRAS.425.2049H}, applying the same classification, as in \citep[][]{2007MNRAS.375..489H} with comparable results \citep[][]{2014MNRAS.445..988N}. Certainly, beyond the linear velocity-density relation, the velocity shear carries complementary information to the tidal field tensor  (see Sec.~\ref{sec:bias}).
The tidal field has also been used to compute the spine of the cosmic web \citep[][]{2008MNRAS.383.1655S,2010ApJ...723..364A,2018JCAP...05..027F}.
Another perspective to the cosmic web is based on  folding of phase-space  \citep[][]{2008MNRAS.385..236V,2009MNRAS.392..281W,2011MNRAS.413.1419V,2012PhRvD..85h3005S,2012MNRAS.427...61A,2012ApJ...754..126F}. The dark matter distribution can be regarded as a continuous field, and gravitational collapse as foldings of phase-space, every time shell crossing occurs. In this framework, voids are regions in which shell crossing has not happened yet, and according to the number of shell-crossings, or the number of axis across which shell-crossing happens, the different cosmic web structures can be identified. As shown in a series of recent works, the regions of  shell-crossing, i.e., the caustics or catastrophies, can be computed from the eigenvalues and eigenvectors of the tidal field tensor \citep[][]{1982GApFD..20..111A,2014MNRAS.437.3442H,2018JCAP...05..027F}.

\subsection{The halo bias picture}
\label{sec:bias}

\noindent 
On the other hand, the bias of galaxies, galaxy clusters, and halos with respect to (w.r.t.) the underlying dark matter field has been studied since long in galaxy surveys and numerical simulations \citep[e.g.][]{1974ApJ...187..425P,1978MNRAS.183..341W,1984ApJ...284L...9K,1985MNRAS.217..805P,1986ApJ...304...15B,1988ApJ...333...21S,1994ApJ...426...23F,1996MNRAS.282..347M,1999MNRAS.308..119S,1999ApJ...520..437K,2000MNRAS.318..203S,2002PhR...372....1C,2010ApJ...724..878T,2011A&A...525A..98V}.
It has been established that the bias relation is non-linear, stochastic, and scale-dependent \citep[e.g.][]{1980lssu.book.....P,1984ApJ...276...13S,1993ApJ...413..447F,1993ApJ...417..415C,1994ApJ...427...51B,1995MNRAS.274..213S,1999ApJ...520...24D,1999MNRAS.304..767S,2001MNRAS.320..289S,2002MNRAS.333..730C,2007PhRvD..75f3512S,2010PhRvD..82j3529D,2012PhRvD..85h3002S,2013PhRvD..88h3507B,2013PhRvD..88b3515S,2014MNRAS.439L..21K,2014MNRAS.441..646N,2015MNRAS.450.1486A}.
Furthermore, it also depends on the history each volume element of the Universe has experienced \citep[e.g.][]{2007MNRAS.377L...5G,2008MNRAS.387..921A,2017JCAP...03..059L,2018JCAP...10..012C,2019MNRAS.484.1133C}. 
This so-called assembly bias can be expressed in terms of the tidal field tensor,
of the velocity field, and of some sort of short-range density dependence,  which altogether also account for non-local bias \citep[][]{1999ApJ...525..543M,2006PhRvD..74j3512M,2008PhRvD..77f3530M,2012PhRvD..85h3002S,2012PhRvD..85h3509C,2012PhRvD..86h3540B,2013PhRvD..88h3507B,2013PhRvD..87h3002S,2013JCAP...08..037P,2013PhRvD..88b3515S,2014JCAP...05..022P,2014PhRvD..90l3522S,2014JCAP...08..056A,2015JCAP...02..013S,2015JCAP...09..029A,2015JCAP...07..030M,2015JCAP...11..007S,2017MNRAS.472.3959M,2018JCAP...09..008L,2018JCAP...07..029A,2018JCAP...01..053M}. 

\noindent 

We investigate  an effective bias model at the Eulerian field level
 \citep[][]{2019PhRvD.100d3514S} restricted to the following dependencies between the halo field  $\delta_{\rm h}$ and the dark matter (DM) gravitational potential:
\be
\label{eq:fbias}
\delta_{\rm h}(\vec r) \curvearrowleft     P\left(\delta_{\rm h}(\vec r) | \nabla^2\phi(\vec r), \partial_i\partial_j\phi(\vec r)\right)=P\left(\delta_{\rm h}|\delta, \mathcal T\right)\,.
\ee

This relation is not truncated to any order, but effectively corresponds to resummed perturbation theories \citep[see e.g.][]{2011MNRAS.416.1703E},  including the infinite non-linear expansion of $\delta$ and the  tidal anisotropies to higher orders, as we show in this work (see also \S\ref{sec:bias} and \citet[][]{2020MNRAS.493..586P}).

This work is especially inspired by \citet[][]{2009JCAP...08..020M} and \citet[][]{2012PhRvD..85h3509C}. In these works, the halo-bias terms are constructed from the tidal field tensor $\mathcal{T}$ and the velocity shear $\Sigma_{ij}$.  For a modern and accurate description of both the two and three-point statistics of biased tracers of the large-scale structure based on perturbation theory see \citet{2021PhRvD.103l3550E} and other recent works such as \citet{2018JCAP...09..008L}. As mentioned before, in the linear regime, and in the absence of vorticity of the velocity field   \citep[washed away due to the cosmological expansion, see e.g. ][]{2002PhR...367....1B} the velocity shear and the tidal field tensor coincide.

The tidal field can be identified among the  precursors of the so-called assembly bias \citep[e.g.][]{2007MNRAS.377L...5G,2008MNRAS.387..921A,2017JCAP...03..059L,2018JCAP...10..012C}. Indeed, tidal torque theory   \citep[][]{1970Afz.....6..581D,1979MNRAS.188..273B,1984ApJ...286...38W,1987ApJ...319..575B,1988ApJ...329....8H,1988MNRAS.232..339H,1996MNRAS.282..436C,1996MNRAS.282..455C,2002MNRAS.332..325P,2002MNRAS.332..339P,2009JCAP...02..023L,2017PhRvD..95f3527C}
predicts an angular momentum of proto-halos of the form \citep[][]{1984ApJ...286...38W} $L_{i }\propto\epsilon^{\rm LC}_{ijk}\,\mathcal{T}_{j \ell} \mathcal{I}_{\ell k}$, where $\mathcal I_{ij}\propto\int_V{\rm d}^{3}q\,  q_i q_j$ denotes the inertia tensor of the mass contained in the volume $V$ (with $\epsilon^{\rm LC}_{kij}$ being the Levi-Civita tensor). This represents the seeds of the spin of dark mater halos, as well as anisotropies in the environment  \citep[][]{2019MNRAS.489.2977R} inducing to different clustering signals of present-day haloes. Furthermore, the late cosmological times, this can induce curl in the velocity field $\omega_{i}\propto \epsilon^{\rm LC}_{ijk}\partial_j v_k$, with non negligible impact in the formation of dark matter halos \citep[][]{1999A&A...343..663P,2015MNRAS.454.3920H}.

As pointed out by \citep[][]{2009JCAP...08..020M}, short-range non-local bias at a given scale $R$ can be modelled by a series of higher order derivative terms like $R^n\nabla^n\delta$ (such as $n=2$) \citep[see also][]{2020MNRAS.492.1614W}. Cosmic voids can be described, as such a case of short range non-local bias, where the density field curvature changes on relatively small scales \citep[][]{2017PhRvD..95f3528C}.
Also, the local tidal environment has been studied with a tensor of the form: $\partial_i\partial_j\delta$ \citep[][]{1988MNRAS.232..339H}. Let us summarise all these short range non-local bias terms by $\Gamma^l_{ij}=\partial^l_i\partial^l_j\delta$ for $i=j$, $i\neq j$, and $l\in\mathbb{N}$.  The mixed terms with $i\neq j$ have been considered in many other works. They enter at higher orders than the $\nabla^2\delta$ term, but are relevant for the bi-spectrum \citep[see for example,][]{2018PhR...733....1D,2018JCAP...02..058N,2019PhRvD..99l3514E,2020JCAP...10..059F,2021PhRvD.103l3550E}. There are also approaches that measure the various perturbative galaxy bias ingredients from simulations, which means they also contain all non-linearities \citep[][]{2021MNRAS.505.1422K,2021arXiv210112187Z}.

In addition, the halo distribution represents a discrete realisation of the expected number counts of objects per volume element. This causes a stochastic uncertainty component, which can be modelled by an additive shot-noise term in the power spectrum measurements \citep[][]{1994ApJ...426...23F}. In large scale structure analysis this uncertainty can be modelled by a white noise term. We refer to the general stochastic bias component as $\epsilon$.

In summary, the halo bias model would have the following dependencies:
\be
\delta_h(\vec r)=F(\delta,\mathcal{T},\Gamma,\Sigma,\omega,\epsilon)\;.
\ee

We  consider each dependency grouped in separated dependencies, such as the combined local and non-local density dependence.

Since the halo number density is a scalar quantity any non-local bias term needs to be  some kind of contraction  of a tensor. In particular, let us consider the series of higher order non-local bias terms constructed based upon $\mathcal T_{ij}$, and rewrite the convenient trace-less tensor as
\be
\label{eq:s2}
s_{ij}\left(\vec r\right)\equiv
\mathcal{T}_{ij} - \frac{1}{3} \delta^{\rm K}_{ij}
\delta\left(\vec r\right)
\,.
\ee
The corresponding second-order and third order bias terms are respectively  given  by the following contractions: 
$s^2\equiv s_{ij}s_{ji}$, 
$s^3\equiv  s_{ij} s_{kj} s_{ki}$.
We consider a general local and non-local functional dependence on the density field $F_{\delta,\mathcal T}(\delta)\equiv F_\delta(\delta,\mathcal{T}_{ij})$.

We note that the velocity terms considered in \citep[][]{2009JCAP...08..020M} correspond to $F_{\rm shear}(\vec v|_{\rm curl-free})$ for curl-free fields, there is an additional shear field for the divergence-free field $F_{\rm shear}(\vec v|_{\rm div-free})$, according to the Helmholtz decomposition \citep[][]{helmhotz59}. Also, the curl of the divergence free velocity field does not vanish and additional terms can be constructed.
 Additionally,  we group the short-range density terms by   $F_{\delta^{\rm SR}}(\partial^l_i\partial^l_j\delta(\vec r))$,
and all the noise terms in
 $F_\epsilon(\epsilon(\vec r))$. 
 
 The resulting Taylor expanded bias dependence following \citet{2009JCAP...08..020M}
 to model the halo over-density in  Eulerian coordinates $\vec r$ is given by:
\begin{widetext}
\ba
\label{eq:bias}
\delta_{\rm h}(\vec r)&=& \underbrace{
\underbrace{
\overbrace{
c_\delta\delta(\vec r)
}^{\mathrm{local}}
}_{\mathrm{first\; order}} 
+
\underbrace{
\overbrace{
\frac{1}{2} c_{\delta^2}\left( \delta^2(\vec r)-\langle\delta^2\rangle\right)
}^{\mathrm{local}} 
+
\overbrace{
\frac{1}{2} c_{s^2} \left(s^2(\vec r)-\frac{2}{3}\langle\delta^2\rangle\right)
}^{\mathrm{non-local}}
}_{\mathrm{second\; order}}  
+ 
\underbrace{
\overbrace{
\frac{1}{3!} c_{\delta^3} \delta^3(\vec r)
}^{\mathrm{local}} 
+
\overbrace{
\frac{1}{2}c_{\delta s^2} \delta(\vec r) s^2(\vec r) +
\frac{1}{3!} c_{s^3} s^3(\vec r)
}^{\mathrm{non-local}}
}_{\mathrm{third\; order}}
+\overbrace{
\underbrace{
\mathcal{O}(F_{\delta,\mathcal T}(\delta)|^4)
}_{\mathrm{fourth\; order+}}}^{\mathrm{local\;\&\;non-local}} 
}_{\mathrm{curl-free\;\&\;\theta=\delta\;terms}}
\nonumber\\
&&
\overbrace{
+
\underbrace{
\underbrace{
F_{\delta^{\rm SR}}(\partial^l_i\partial^l_j\delta(\vec r))
}_{\mathrm{short\;range\;\theta=\delta\;terms\;\it  l\in\mathbb{N}}}
}_{\mathrm{first\;order+}}
+ 
\underbrace{
\underbrace{
F_{\rm shear}(\vec v(\vec r)|_{
\rm curl-free
})
}_{\mathrm{\theta\neq\delta\;terms}}
+
\underbrace{
F_{\rm shear}(\vec v(\vec r)|_{
\rm div-free
})
+ 
F_{\rm curl}(\vec v(\vec r))
}_{\mathrm{vorticity\;\theta=\delta\;\&\;\theta\neq\delta\;terms}}
}_{\mathrm{third\;order+}}
}^{\mathrm{non-local}}
+
\overbrace{
\underbrace{
\underbrace{
F_\epsilon(\epsilon(\vec r))
}_{\mathrm{noise\;terms}} 
}_{\mathrm{first\;order+}}
}^{\mathrm{local\;\&\;non-local}} \,,
\ea
\end{widetext}
where $\delta$ stands for the dark matter field, the $c$s are some bias factors, the $s^2$ and $s^3$ are non-local bias terms derived from the tidal field tensor, and $\theta$ is the weighted  velocity divergence. We neglect in this study the velocity terms and  the  short range terms, which are  typically  given for $l=1$ and $i=j$ \citep[][]{2017PhRvD..95f3528C,2020MNRAS.492.1614W}, although also $i\neq j$ has been considered \citep[][]{1988MNRAS.232..339H}.
It is important to stress that in contrast to the majority of previous  works \citep[see e.g.][]{1986ApJ...304...15B,2002MNRAS.329...61S,2009JCAP...08..020M,2012PhRvD..85h3509C}, we consider the gravitational potential from the cosmologically evolved non-linear density field  defined on a few Mpc scales.  At such scales the divergence-free component makes up only a few percent of the  curl-free  component,  and becomes increasingly important towards sub-Mpc scales \citep[][]{2012MNRAS.425.2422K}. Thus, we can assume  approximately  curl-free local density terms, and the first row  can be  identified with the tidal shear for curl-free fields with $\theta=\delta$: $F_{\delta,\mathcal T}(\delta)\simeq F_{\rm shear}(\vec v(\vec r)|_{
\rm curl-free
}^{\theta=\delta})$.

 We are interested in practical applications to efficiently generate mock galaxy catalogues saving the expensive costs of running $N$-body simulations. For this reason, we can, a priori, not assume to have such a precise velocity field to be able to consider the case for which $\theta\neq\delta$. However, we can assume that machine learning algorithms can efficiently deliver such precise velocity fields \citep[see e.g.][]{2021ApJ...913....2W}. We will, hence, also consider the velocity shear as in the pioneering work of \citet[][]{1996Natur.380..603B} to describe the halo distribution. Below, we will start with the tidal field tensor, but the same analysis will be shown for the velocity shear tensor in \S\ref{sec:shear}.

\subsection{Unified cosmic web and halo bias picture}
\label{sec:unified}

\noindent 
The gravitational potential is the key ingredient in both the development of the cosmic web and the bias relation between the halo field and the underlying dark matter density field. Therefore, following \citep[][]{2002MNRAS.329...61S,2012PhRvD..85h3509C}, we analyse the gravitational deformation tensor in terms of its invariants \citep[see also][]{2018JCAP...01..053M}.  
The corresponding eigenvalues of the symmetric tidal tensor $\mathcal T$ are computed by solving its cubic
characteristic polynomial \citep[see e.g.][]{mechanics}:
$\det(\lambda \mathbb{1} - \mathcal T) = \lambda^3-I_1\lambda^2+I_2\lambda-I_3 = 0$, 
with invariants: 
\ba
\label{eq:princi}
I_1&\equiv&{\rm tr}(\mathcal T)=\lambda_1+\lambda_2+\lambda_3\equiv\delta\;{\rm  with}\;\lambda_1\ge\lambda_2\geq\lambda_3\nonumber\,,\\ I_2&\equiv&\frac{1}{2}({\rm tr}^2(\mathcal T)-{\rm tr}(\mathcal T^2))=\lambda_1\lambda_2+\lambda_1\lambda_3+\lambda_2\lambda_3\nonumber\,,\\  I_3&\equiv&\det(\mathcal T)=\lambda_1\lambda_2\lambda_3\nonumber\,,\\
I_4&\equiv&{\rm tr}(\mathcal T^{\rm t} \mathcal T)=\lambda_1^2+\lambda_2^2+\lambda_3^2=I_1^2-2I_2\nonumber\,,\\ 
I_5&\equiv&{\rm tr}(\mathcal T^{\rm t}\mathcal T^{\rm t} \mathcal T)=\lambda_1^3+\lambda_2^3+\lambda_3^3\nonumber\\ &=& I_1^3-3I_1I_2+3I_3=I_1I_4-I_1I_2+3I_3\,.
\ea

\noindent Defining $9\,\alpha\equiv I_1^2-3I_2=I_4-I_2$, $\beta\equiv(-9I_1I_2 +27I_3+2I_1^3)/54$, and $\Theta\equiv \cos^{-1}(\beta/\sqrt{\alpha^3})/3$, it is possible to formulate the eigenvalues as \citep[][]{weisstein02,press2007numerical,Nickalls1993ANA}:
 $\lambda_1=I_1/3+2\sqrt{\alpha}\cos({\Theta})$,  $\lambda_2=I_1/3+2\sqrt{\alpha}\cos({\Theta-2\pi/3})$, and $\lambda_3=I_1/3+2\sqrt{\alpha}\cos({\Theta+2\pi/3})$.
From this,  the moments can be constructed \citep[][]{Pierpaoli1996TowardAQ}:
 $\mu_1=\langle\lambda_i\rangle=I_1/3$, 
$\mu_2=\langle(\lambda_i-\mu_1)^2\rangle=2\,\alpha$, 
 and $\mu_3=\langle(\lambda_i-\mu_1)^3\rangle=2\,\beta$.
 The second order non-local bias $s^2$ term \citep[see][]{2009JCAP...08..020M} is also related to $\alpha$  \citep[][]{2018MNRAS.476.3631P}:  $s^2=6\,\alpha$. Hence, $s^2=2/3\,(I_4-I_2)=I_4-1/3I_1^2$. In the same way $s^3$ is related to $\mu_3$:  $s^3=3\,\mu_3=6\,\beta=-I_1I_2+3I_3+2/9I_1^3=1/3(I_5-1/3I_1^3)+2I_3$.  The difference among the eigenvalues induces a non vanishing $\alpha$,  eventually also causing an ellipticity $e\equiv(\lambda_1-\lambda_3)/(2\delta)$ and a  prolatness $p\equiv(\lambda_1+\lambda_3-2\lambda_2)/(2\delta)$, which require knowledge on the first three invariants  \citep[][]{2012PhRvD..85h3509C}: $I_2=I_1^2/3\,(1-(3\,e^2+p^2))$ and $I_3=I_1^3/27\,(1-p)[(1+p)^2-9\,e^2]$.
Following \citep[][]{2012PhRvD..85h3509C} we can find the relation between the halo bias terms from \citet[][]{2009JCAP...08..020M} and the invariants, as generating functions), respectively (see \S\ref{sec:bias}). 
From Eq.~(\ref{eq:princi}) it is clear that each of the eigenvalues can be expressed as a function of the first three invariants. As a consequence, the classical cosmic web   classification, dubbed T-web, (based on combinations of conditions on the eigenvalues to obtain four different cosmic web types: $t_{\rm cw}$:=\{knot, filament, sheet, void\}, see \citep[][]{2007MNRAS.375..489H}) has an equivalent formulation in terms of conditions on the invariants, as we show in the appendix \ref{sec:cwinv}.

\section{Gravity calculations with numerical simulations}
\label{sec:nbody}

 We rely in this work on the Minerva suite  \citep{2016MNRAS.457.1577G}, which consist on a set of $300$ $N$-body dark matter only simulations, each embedded in a cubic box of $L_{\rm box}=1.5\,h^{-1}$ Gpc side using 1000$^3$ particles. Dark matter haloes are identified with a standard Friends-of-Friends (FoF) algorithm at redshift $z = 1$, and subjected to an unbinding procedure \citep{2001MNRAS.328..726S}, in which particles with positive total energy are removed and halos artificially
linked by FoF are separated. The minimum halo mass is $\sim 2.7\times 10^{12}\, h^{-1}\,M_{\odot}$. 
The advantage of these  simulations  is that they have been thoroughly studied in the  different summary statistics   \citep[][]{2019MNRAS.485.2806B,2020MNRAS.491.2565B}, and thus  constitute an ideal reference set. 

This population of halos is particularly sensitive to a well modelling of non-local bias in the three-point statistics.
Luminous Red Galaxies, correspond to an effective higher mass cut ($>10^{13} h^{-1}\,M_{\odot}$), and thus the corresponding halos better trace the peaks of the density field. These can be described with simpler bias models \citep[][]{2015MNRAS.450.1836K,2016MNRAS.456.4156K}. Although, in these studies the bi-spectra were not reproduced to the accuracy presented in this work. On the other hand considering very low mass halos ($>10^8 h^{-1}\,M_{\odot}$) brings the halo field closer to the dark matter field, and no strong evidence for non-local bias terms was found \citep[][]{2020MNRAS.493..586P}. 
The rather intermediate halo mass cut considered in this study constitutes the most challenging one, and hitherto no well modelling of the three-point statistics has been found on the scales relevant to baryon acoustic oscillations, or redshift space distortions \citep[see][]{2020MNRAS.491.2565B}. A good handling on halo masses above $10^{11}$ and below  $10^{13} h^{-1}\,M_{\odot}$ is critical for the upcoming surveys focusing on emission line galaxies, such as \citep[][]{DESI,Euclid,jpas,4most}.

Since, we aim at studying the relation between the formation of compact  small scale objects and the large scale structure, we define a regular mesh (with cell center positions at $\vec r$) of 500$^3$ cells, which implies a 3 $h^{-1}$ Mpc cell side resolution.
We apply nearest-grid-point to the halo catalogue and to the dark matter particles, to produce halo number  counts per cell $N_{\rm h}(\vec r)$ and the dark matter field $\delta(\vec r)$, respectively.

The cell size has been chosen to be large enough that there is a considerable computing gain, but small enough that it is still below the scale of typical displacements to avoid an additional smoothing which could have an impact on Baryon Acoustic Oscillations or Redshift Space Distortions. A deeper study in this direction should be performed to investigate whether even smaller resolution meshes can be safely considered at the expense of requiring more computing resources.

\begin{figure*}
\vspace{-.4cm}
\begin{tabular}{c}
\hspace{-0.5cm}
\includegraphics[width=1.05\textwidth]{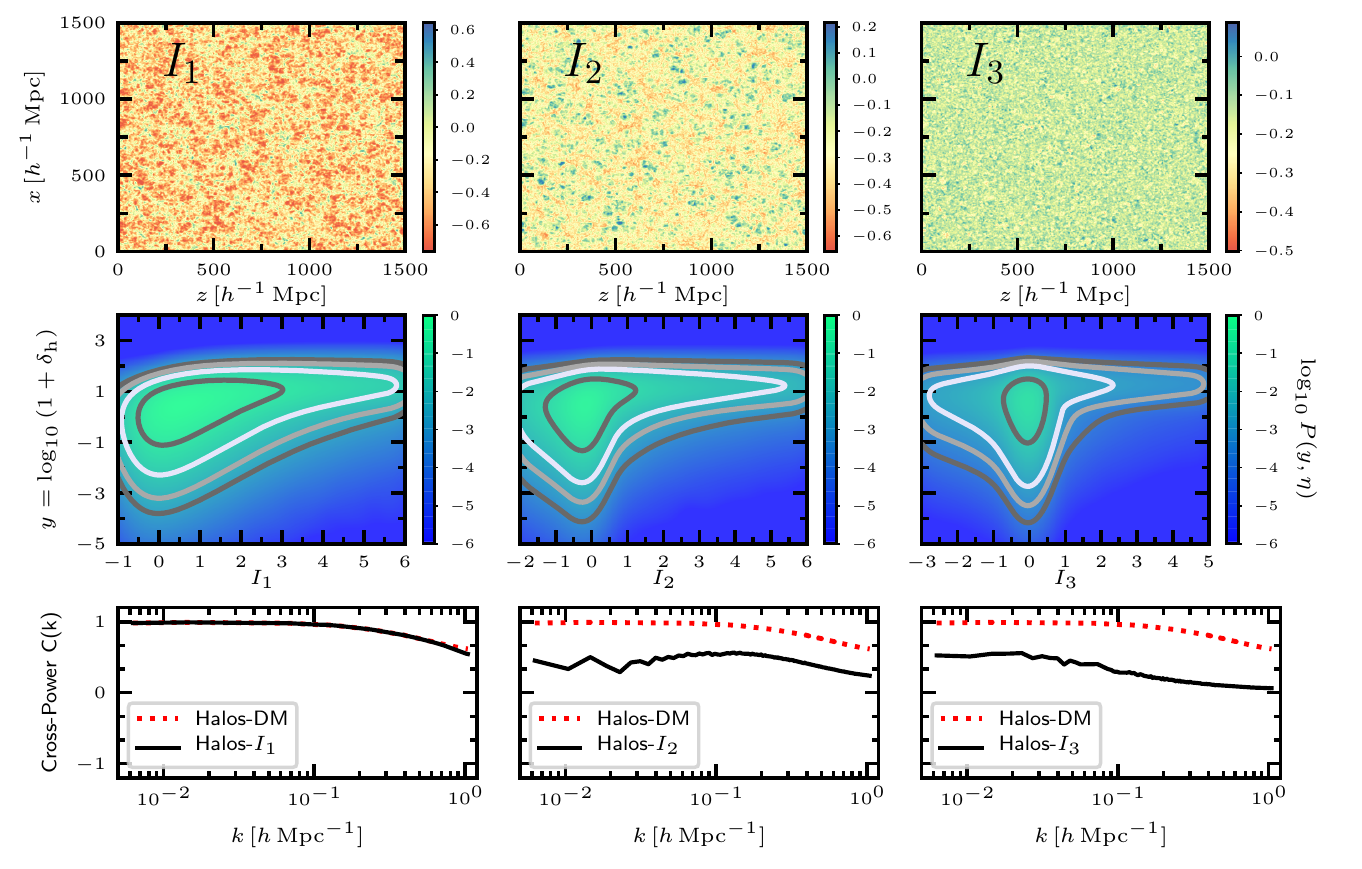}
\vspace{-.3cm}
\\
\hspace{-0.5cm}
\includegraphics[width=1.05\textwidth]{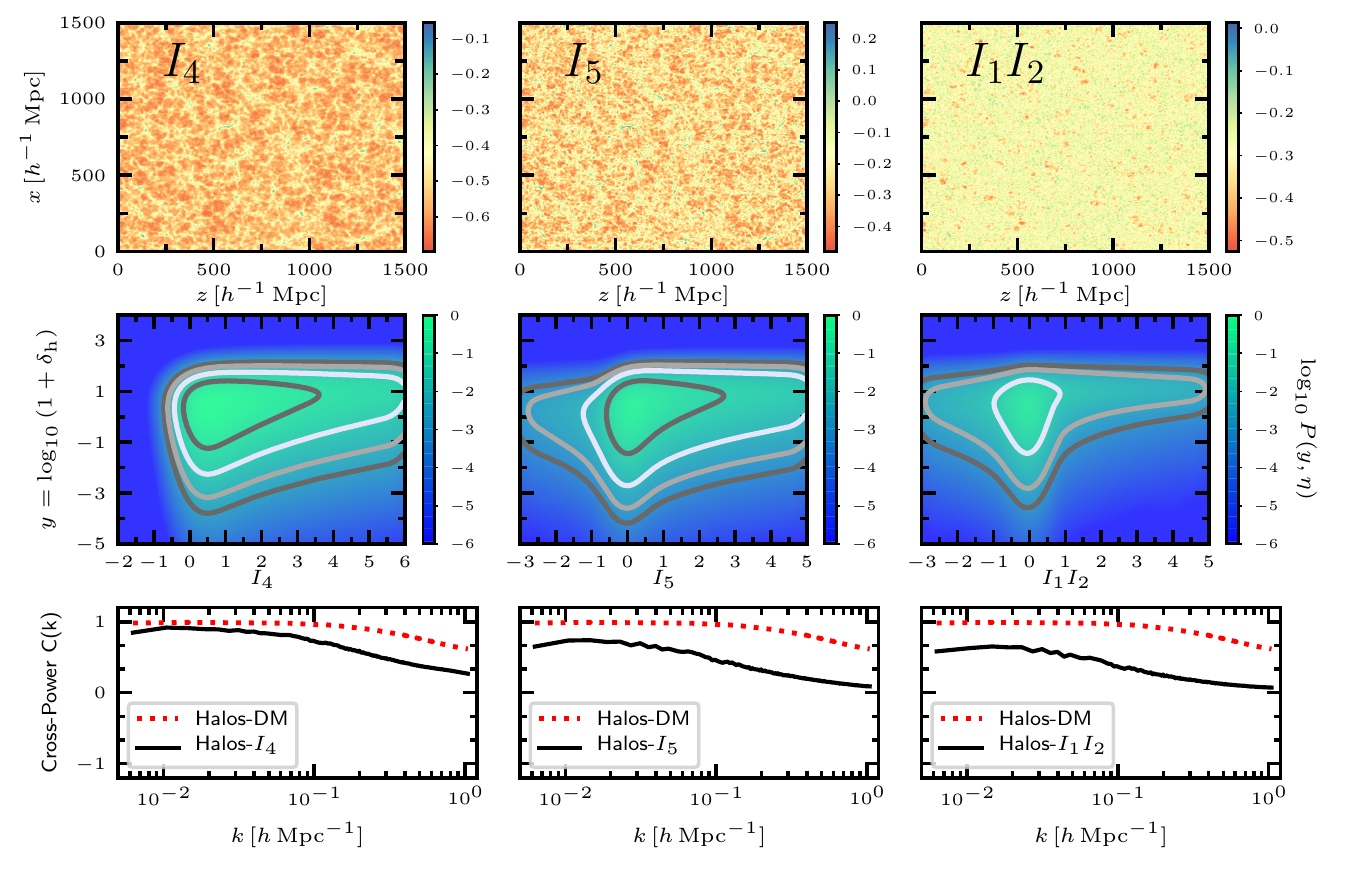}
\end{tabular}
\vspace{-.7cm}
\caption{Generating functions $\{I_1,I_2,I_3,I_4,I_5,I_1I_2\}$ and their relation to the halo field extracted from $N$-body simulations. Upper panels show slices through the 1.5 $h^{-1}$ Gpc side volume for each variable. The second row shows the smoothed bias relation with respect to the halo number over-density. The contours in each panel denote the region containing 65, 95 and 99\% of the total number of classified cells. The lower panels show the corresponding  cross power spectra.} 
\label{fig:bias}
\end{figure*}

\begin{figure*}
\hspace{-0.5cm}
\includegraphics[width=1.05\textwidth]{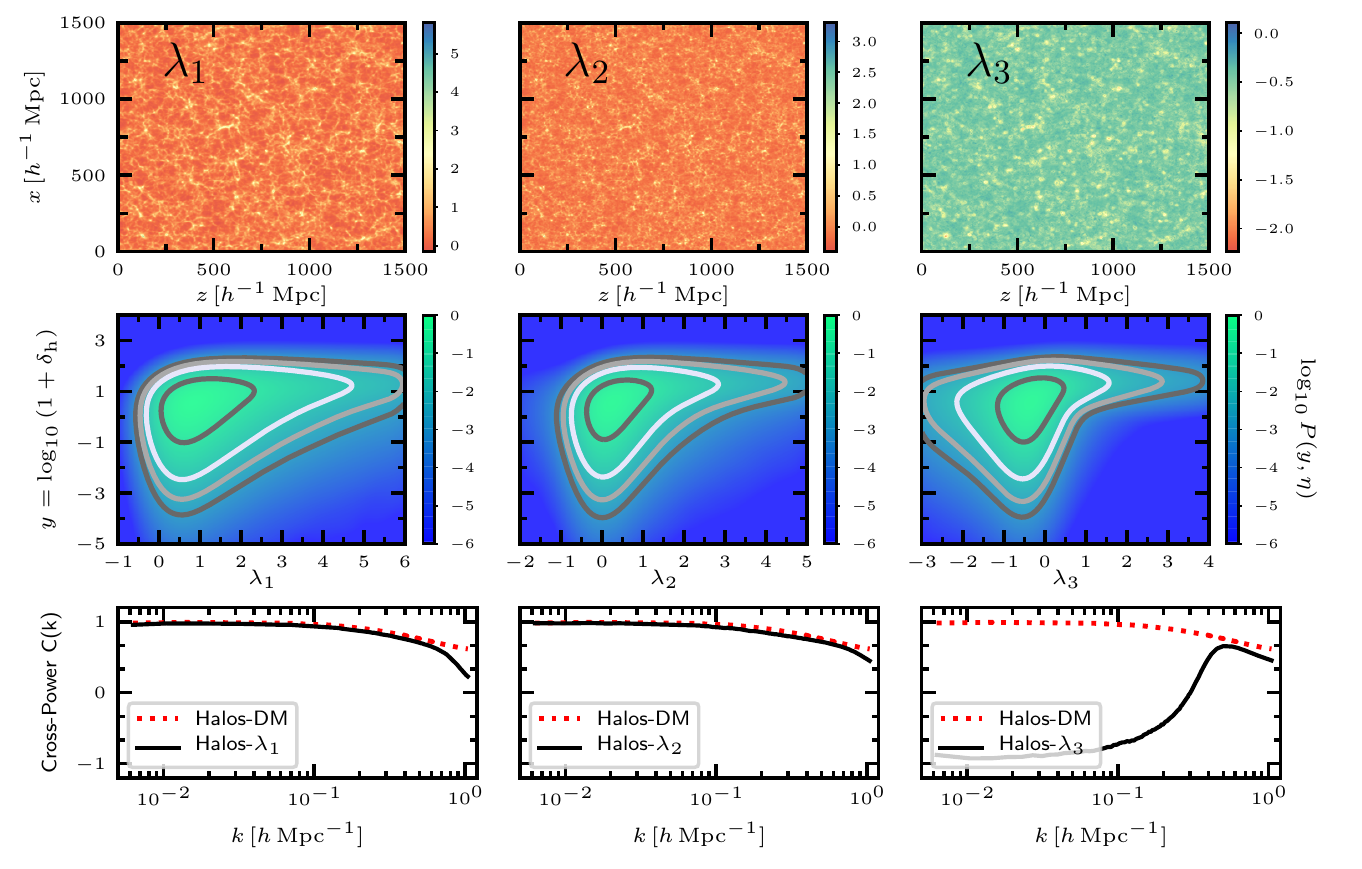}
\vspace{-0.5cm}
\caption{
Generating functions $\{\lambda_1,\lambda_2,\lambda_3\}$ and their relation to the halo field extracted from $N$-body simulations. Upper panels show slices through the 1.5 $h^{-1}$ Gpc side volume for each variable. The second row shows the smoothed bias relation with respect to the halo number over-density. The contours in each panel denote the region containing 65, 95, 97 and 99\% of the total number of classified cells. The lower panels show the corresponding  cross power spectra.} 
\label{fig:bias2}
\end{figure*}

\section{BAM: a machine learning method to combine generating functions}
\label{sec:bam}

The \texttt{BAM} method represents a  physically motivated supervised machine learning algorithm in which the cost function based on the power spectrum of the respective targeted variables is minimized with non-linear and non-local isotropic kernels and anisotropic explicit bias dependencies. The kernels preserve dimensionality and are iteratively extracted from the reference simulation.  

The bi-spectrum is not explicitly used in the calibration process. Nevertheless, using an accurate dark matter field above a certain scale (a few Mpc) combined with an accurate bias modelling ensures a precise three-point statistics of the tracer distribution \citep[see][]{2015MNRAS.450.1836K}.
As in all previous works using the \texttt{BAM} algorithm, we describe  the  joint  probability  distribution of the number of halos and all considered properties \citep[see][]{2020MNRAS.491.2565B}.

In this work, we use in the calibration process a down-sampled version of the initial conditions and the final catalogue from the reference simulation.

Considering the general problem of a quantity $q$ determined by the functional dependence on a series of variables, which are generating functions $\{\eta_1,\eta_2,\dots\}$:  $q=F(\eta_1,\eta_2,\dots)$, with each variable having a  potentially non-linear and non-local relation w.r.t. $q$: $q=F_i(\eta_i)$ $\forall i$.
Let us further assume that we have a method to express the dependence, as  combinations of non-linear expansions of each variable: 
\ba
\label{eq:qvdep}
q&=&c_1\,F_1(\eta_1)+c_2\,F_2(\eta_2)+c_{12}\,F_{12}(\eta_1\eta_2)+\cdots\\
&=&c_1(a_1\,\eta_1+a_2\,\eta_1^2+\cdots)+c_2\,(b_1\,\eta_2+b_2\,\eta_2^2+\cdots)+\cdots \nonumber\,.
\ea
If there is then a relationship between the different variables: $v_i=f_{i}(\{\eta_j: j\neq i\})$, the question arises of which is the minimum number of variables, which fully constrains $q$.
 In practice, it is difficult to  implicitly include cross terms such as $F_{12}(\eta_1\eta_2)$. The reason is that a particular finite binning must be used to describe each quantity $\eta_i$. Hence, cross products are usually not well modelled by the original binning of each component. Therefore, a higher accuracy is achieved when those cross terms are explicitly included.

\subsection{Technical details of the BAM algorithm}

Our particular science case consists of reproducing the halo number counts per cell $N_{\rm h}$ on a mesh of an $N$-body simulation as a function of variables \citep[see e.g.][]{2014MNRAS.439L..21K}.

It is important to note, that the \texttt{BAM} code used in this work does not truncate the bias relation to any order, but effectively corresponds to resummed perturbation theories \citep[see e.g.][]{2011MNRAS.416.1703E}, including the infinite non-linear expansion of $\delta$ and the  tidal anisotropies to higher orders, as we show in this work.  This has  already been studied to some extent (depending on $\delta$ and  $\delta$+T-web)  in \citet{2019MNRAS.483L..58B,2020MNRAS.493..586P}.

The \texttt{BAM} algorithm takes a set of binned variables $\{\eta\}$ and measures from the reference simulation the probability distribution $P(N_{{\rm h}}|\{\eta\})$ for halo number counts per cell $N_{\rm h}$ conditional to the set $\{\eta\}$.
This conditional probability distribution function represents the joint stochastic and deterministic halo bias expressed in terms of the chosen set of variables.

In other words, the noise terms encoded in Eq.~(\ref{eq:bias}) through $F_\epsilon(\epsilon(\vec r))$ are all effectively included in our study. \texttt{BAM} then goes through each cell $i$ in the mesh looking up the joint set of (binned) variables $\{\eta\}_i$ corresponding to that cell, and randomly selects a halo number count $N_{{\rm h}}$ from the available probabilistic relation $P(N_{{\rm h}}|\{\eta\})$. 
This  relation however, depends on the definition of the dark matter density field $\delta$ on the mesh, from which all invariants are derived.
But the definition of $\delta$  is arbitrary, and does not coincide with the way in which the dark matter distribution was used to define the halos \citep{2001MNRAS.328..726S}. There is an effective kernel relating the mass assignment scheme used to define the density field on the mesh and the  halo finder used to define the halos. In addition, all the missing contributions to the bias (for instance short range  non-local bias) can potentially have an impact on the power spectrum. 
For all these reasons, \texttt{BAM} applies a kernel to the dark matter density.
This kernel is automatically determined through  a Markov Chain Monte Carlo rejection sampling algorithm.
The kernel  purpose is meant to  reproduce the power spectrum of the reference one. However, depending on the bias model a higher or lower accuracy is achieved.  
But the (reduced) three-point statistics  corresponding to the halo realisation produced by \texttt{BAM} is only constrained by the chosen bias model. Therefore, we focus on this statistics to study the different bias models listed above.
Some crucial improvements in the treatment of the variables, which allow us to use a low number of bins to accurately describe them are presented in \citet{HYDROBAM1}.
We  initially use  in all our runs 300 bins for the over-density $\delta$ and 700 bins for the rest of variables, with the exception of the T-web which uses four bins for the cosmic web type  and additional 200 bins for the classification of knots \citep[see][]{2015MNRAS.451.4266Z}. We have checked for the I-web case only, that the results do not qualitatively change using a far lower resolution of bins of \{200,100,100\}. An investigation of potential over-fitting is presented in \citet[][]{HYDROBAM1}. An investigation of how to use multiple  reference simulations -thereby effectively increasing the reference volume- to produce accurate covariance matrices is presented in a forthcoming publication (Balaguera-Antolínez et al., in prep.).

\subsection{Halo bias calculations with invariants}

\label{sec:invariants}

A bias model including the non-linear dependence on $\delta$, $I_4$, and $I_5$ includes all the terms in the first row of Eq.~(\ref{eq:bias})  up to third order, with the exception of the term involving $\delta s^2$  and $s^3$ (as it does not include $I_3$).  
 This accounts for the second order anisotropic clustering. 
 The bias contours and the cross-power spectra between the halo field, the invariants and the eigenvalues  are shown in  Figs.~\ref{fig:bias} and Fig.~\ref{fig:bias2}, respectively.
The fractional and relative anisotropy (FA and RA, respectively), commonly used quantities in medical imaging \citep[see e.g.][]{neuro}, are related to $\alpha$: ${\rm FA}=\sqrt{3/2\,(\sum_i(\lambda_i-\mu_1)^2)/I_4}=\sqrt{3\mu_2/(2I_4)}=\sqrt{3\,\alpha/I_4}$ and ${\rm RA}=\sqrt{1/6\,\mu_2/\mu_1^2}=\sqrt{3\,\alpha}/I_1$.
The latter being similar to the halo-centric anisotropy \citep[][]{2019MNRAS.489.2977R}: $\alpha'\equiv\sqrt{\alpha}/(1+\delta)$.
From the main invariant $I_5$ we see that the term $I_1I_2$ is fully specified, if in addition to $I_5$ and $I_1$ (through $\delta$), the third principal invariant $I_3$ is specified (see Eq.~(\ref{eq:princi})).
On the other hand, from the definition of $I_4$, we can see that  $I_4$ and $I_1^2$ (through $\delta$) generate $I_2$.
Reversing the argument, by specifying $\delta$ and $I_2$ we automatically fix $I_4$.
In this way, by describing the non-linear relation between the halo density and 
$\{\delta(I_1),I_2,I_3,I_4,I_5\}$ 
(of which $I_4$ is redundant), all the terms from the non-linear local and non-local density bias relation $F_{\delta^{\rm NL}}(\delta)$ up to (at least) third order are fully specified.
In fact, even the common  anisotropic functional dependencies are  described  with the invariants.
Since specifying \{$\delta$,$I_2$,$I_3$,$I_5$\} corresponds to four equations involving three eigenvalues (see Eq.~(\ref{eq:princi})), each eigenvalue is constrained.
In this way, the ellipticity $e$  and the prolatness $p$, as defined  in Sect. \ref{sec:cw},  are fixed. In fact, it is clear now, that $e$ and $p$ involve the invariants $I_1$, $I_2$, and $I_3$, corresponding thus to non-local bias terms which are not complete at  third order in Eq.~(\ref{eq:princi}), since they do not involve $I_5$.

We conclude that the invariants \{$I_1,I_2,I_3,I_4,I_5$\} generate  the following terms in the halo bias relation to third order (Eq.~\ref{eq:bias}): $\delta,\delta^2,\delta^3,s^2(\alpha),s^3,\delta\,s^2,e,p$.

The invariant $I_4$ is generated by $I_1^2$ and $I_2$. Hence, a method able to make combinations of the type described by Eq.~(\ref{eq:qvdep}) requires only $I_1$ and $I_2$ to automatically also include $I_4$.
$I_5$ on the other hand requires $I_1^3$, $I_3$, and $I_1I_2$. We need thus to include $I_3$ together with  $I_1I_2$, which is also required to model $\delta\,s^2$ by $I_1I_4=I_1^3-2I_1I_2$. 

Then the set of variables $\{I_1,I_2,I_3,I_1I_2\}$ is equivalent to  $\{I_1,I_2,I_3,I_4,I_5\}$.
In terms of the statistical moments of the eigenvalues the set $\{I_1,I_2,I_3,I_1I_2\}$ determines $\{I_1,\alpha,\beta\}$, and hence, all moments to third order $\{\mu_1,\mu_2,\mu_3\}$.

We can also replicate the invariants $I_{i}$ with the eigenvalues. A first naive attempt consists of using a model such as $\{I_1,\lambda_1,\lambda_2,\lambda_3\}$. This however, will not include terms  like  $\lambda_i\lambda_j$, nor $\lambda_i^2\lambda_j$ with $i\neq j$. This implies that terms such as $I_2$, $I_3$, $I_1I_2$  cannot  be constructed. It is tempting to construct combined variables such as $\lambda_{ij}=\lambda_i+\lambda_j$ with $i\neq j$ to account for those mixed terms. However, according to Eq.~(\ref{eq:qvdep}) by doing so we would have the same weights for terms which require an independent treatment to construct the various bias terms. 
 One could further extend this model including $I_1I_2$, however at the computational expense of having five variables instead of four to describe all relevant bias terms up to third order neglecting velocity  and short range non-local terms. Nontheless, the model  $\{I_1,\lambda_1,\lambda_2,\lambda_3\}$ constrains powers of $\delta$, and \{$I_4$,$I_5$,$e$,$p$\}. It is thus an interesting model to test the capability of \texttt{BAM} to combine different variables to effectively treat a series of bias terms.

 For computational reasons, it is also interesting to consider the case of $\{\delta,I_2,I_3\}$ (hereafter dubbed I-web or I-web-V when based on the tidal field or velocity shear tensor, respectively),  as compared to the complete invariants set. As shown above, the $I_1I_2$ is well modelled through $I_3$, and hence similar results are expected from including it or not.

 In fact, one needs to realise that $I_1I_2$ is related to $I_3$ through the equation:
\ba
\label{eq:I1I2}
I_1I_2&=&3\,I_3+\sum_{i\neq j}\lambda_i\lambda_j^2\;.
\ea
 From this we see that when all three eigenvalues are similar $I_1I_2\simeq9\,I_3$, implying an  isotropic collapse or expansion, which can happen for cosmic web types such as knots or voids. Moreover, when two eigenvalues are close to each other   $\lambda_1\simeq\lambda_2$ (the same argument is valid for $\lambda_2\simeq\lambda_3$) we have $I_1I_2\simeq3\,I_3+2\,(\lambda_1^3+\lambda_1^2\lambda_3+\lambda_1\lambda_3^2)$  and $I_3\simeq\lambda_1^2\lambda_3$. 
 If we further   demand  all terms to be proportional to $I_3$,  then  $I_1I_2\simeq (3+2c)\,I_3$ with $c$ being a constant factor, which is fulfilled  when $\lambda_1^3+\lambda_1^2\lambda_3+\lambda_1\lambda_3^2\simeq c\,\lambda_1^2\lambda_3$.
 This yields a quadratic equation for $\lambda_3$: $\lambda_3^2+(1-c)\lambda_1\lambda_3+\lambda_1^2\simeq0$. Hence, $\lambda_3\simeq-\lambda_1[1-c\pm\sqrt{(1-c)^2-4}]/2$, which for the particular case of $c=-1$ reduces to $\lambda_3\simeq-\lambda_1$. This is the typical situation of a filament or sheet for which two eigenvalues are similar sharing the same sign, and  the third one has opposite sign. A whole family of valid solutions is given for $c<-1$, which does not force the third eigenvalue to have the same absolute value  as the other two. It is especially when the three eigenvalues are different, that the $I_1I_2$ term adds information, and will need to be investigated further in future work.

 Our study investigates the major contributions up to third order in the halo bias expansion, as they dominate the statistics within the scales we are considering in this work.
 It is important to stress, however,  that \texttt{BAM} does not restrict the dependencies in the variables (generating functions) to any  truncated order.

 The total number of models considered in this study are summarised as follows:
 \begin{itemize}
    \item  Local DM $\delta$: $\{\delta\}$,
    \item  $\delta$+T-web: \{$\delta$, $t_{\rm cw}$=\{knot, filament, sheet, void\}\}  $\sim\{\delta(I_1),\widetilde{s^2}(\widetilde{\alpha}),\widetilde{s^3},\widetilde{e},\widetilde{p}\}$,
    \item PT-web-q:
    $\{\delta,I_4\}$,
    \item PT-web-qa:
    $\{\delta,I_4,s^2(\alpha)\}\sim\{\delta,I_2,I_4\}$\,,
    \item PT-web:
    $\{\delta,I_4,I_5\}$,
    \item  $\lambda$-web:
    $\{\delta,\lambda_{1},\lambda_{2},\lambda_{3}\}\sim\{\delta,I_4,I_5,e,p\}$,
    \item  I-web: $\{\delta,I_{2},I_3\}$    $\sim\{\delta,s^2(\alpha),\widetilde{s^3},\widetilde{\delta s^2},e,p\}$,
    \item  I-web-c: $\{\delta,I_2,I_3,I_1I_2\}\sim\{\delta,I_2,I_3,I_4,I_5\}$ $\sim\{\delta,s^2(\alpha),s^3,\delta s^2,e,p\}$\,,
\end{itemize}
with $\widetilde{x}$ indicating  restricted information on $x$.  As explained in appendix  \ref{sec:cwinv} the T-web has only partial information of the whole parameter space spanned by the invariants (I-web).
There is a  crucial difference between the T-web and the I-web models, which lies in the binning of the variables given by the invariants or the eigenvalues, the same way the density field is binned. The T-web is not only solely based on the invariants $I_1,I_2,I_3$, but also restricts the information to a particular  cosmic web type, and hence corresponds to only four additional bins to the ones used for the density.

\begin{figure*}
\includegraphics[width=\textwidth]{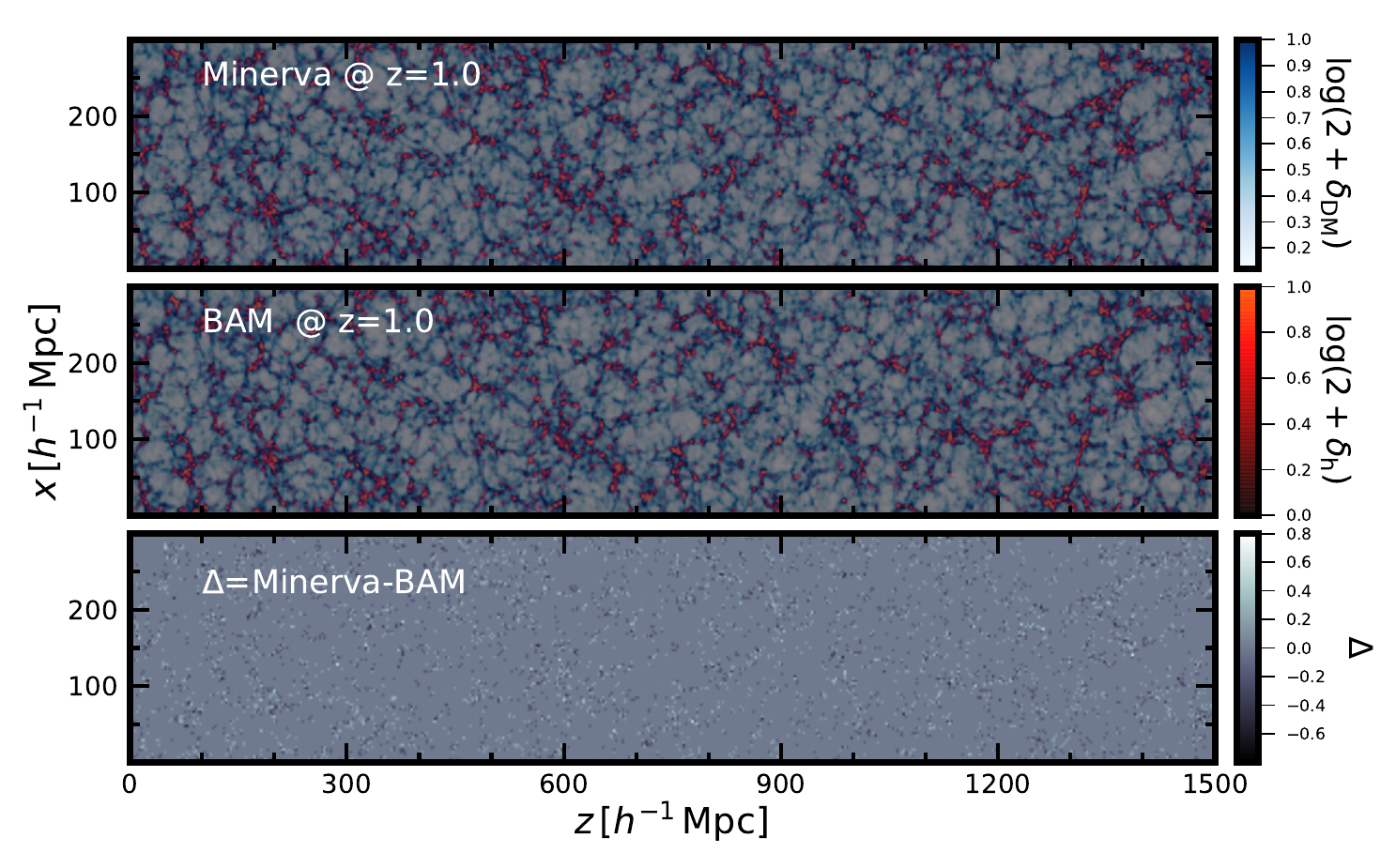}
\caption{ Slices through the simulated cosmic volumes. Top panels: halo over-density field $\delta_{\rm h}$ (red colour range) vs $N$-body dark matter field $\delta$ (blue colour range), top panel: from the $N$-body catalogue (4500 CPU hrs), middle panel: from the \texttt{BAM} mock with the I-web model (0.042 CPU hrs without velocities). Bottom panel: difference between the two halo fields. } 
\label{fig:BAMvsNB}
\end{figure*}

The plots representing the relation between the invariants and the halo density field (such as Figs.~\ref{fig:bias}  or ~\ref{fig:bias2}) show how \texttt{BAM} captures the full non-linear and stochastic relation between each variable and the halo realisation from the reference simulation. This is accurately demonstrated in \citep[][]{2020MNRAS.493..586P}.  
In that work halo populations down to $\gsim  10^{8}\, h^{-1}\,M_{\odot}$ were studied. In that case, the halo population has a very low bias, and in consequence, its two- and three-point statistics are closer to the dark matter field. Hence, simpler bias models, such as $\delta$+T-web succeed to reproduce the summary statistics. However, the non-linear local dark matter density dependence becomes very complex, having an increasing  range of halo number counts towards higher mass resolutions, and simple power law bias models are shown to fail. The study presented in this work, can be thus particularly important for intermediate halo masses, hosting emission line galaxies. 

The invariants of the velocity shear tensor (instead of the tidal field tensor) can be used in the analysis yielding an effective description of further bias terms in Eq.~\ref{eq:bias} (see \S\ref{sec:shear}).

\section{Analysis of gravity calculations}
\label{sec:analysis}

\noindent

In this section, we present the results after  applying the Bias Assignment Method \citep[\texttt{BAM} code:][and   \S\ref{sec:bam}]{2019MNRAS.483L..58B,2020MNRAS.491.2565B,2020MNRAS.493..586P}, which relates the halo number counts per cell in a mesh ($3\hinvm$  cell side resolution in a $1.5\, h^{-1}$ Gpc cubical volume) to a number of dependencies given by Eq.~(\ref{eq:fbias}).

\noindent
 We  focus in this work on the following biasing models: local $\delta$; $\delta$+T-web: \{$\delta$, $t_{\rm cw}$\}; PT-web-q:
    $\{\delta,I_4\}$; PT-web:
    $\{\delta,I_4,I_5\}$; and I-web: $\{\delta,I_{2},I_3\}$.
The invariants and their relation to the halo number over-density $\delta_{\rm h}$ are shown in Fig.~\ref{fig:bias}.  The bias relation and the  cross power spectra of $I_3$ and $I_1I_2$ turn out to have a relatively  similar grained  structure, similar cross power spectra, and bias relations. In fact, $I_1I_2$ is proportional to $I_3$ for knots and voids, when   the three eigenvalues are close to each other, and for filaments and sheets when two eigenvalues are   roughly the same (see \S\ref{sec:cw}).
Hence, we can save in this study the cross term, reaching equal accuracy. It becomes visually clear how $I_1$ and $I_2$ can build $I_4$ (with a $\delta
^2$ term yields $s^2$), and how $I_1$, $I_3$ (and $I_1I_2$) can generate $I_5$ (with a $\delta
^3$ term yields $s^3$). 
The results from  \texttt{BAM} 
 are summarised in Figs.~\ref{fig:PK}- \ref{fig:BK11}.

\noindent  
Several previous works have detected non-local bias in $N$-body  simulations by doing truncated bias expansion  \citep[for quadratic terms  see][]{2012PhRvD..85h3509C,2012PhRvD..86h3540B} and \citep[for cubic terms see][]{2017MNRAS.472.3959M,2018JCAP...07..029A}.  
 We note, that in the present work we manage to confirm such dependencies by reproducing the three-dimensional halo distribution including the phases at small scales, as illustrated in Fig.~\ref{fig:BAMvsNB}. The two halo distributions from the full gravity calculation and the \texttt{BAM} mock using the I-web look very similar. Their tiny differences are a consequence of accounting for the full stochastic nature of the halo bias relation at the cell resolution (see Eq.
~\ref{eq:fbias}).  This is particularly useful, as it permits us to produce accurate synthetic halo catalogues without having to make the expensive computations of $N$-body codes. The \texttt{BAM}-code  is especially  efficient (see computing requirements of the \texttt{PATCHY} code, which are comparable to \texttt{BAM} in contrast to other methods in \citet[][]{2019MNRAS.485.2806B} and \citet[][]{2020MNRAS.493..586P}).

\subsection{Two- and three-point statistics analysis}
\label{sec:bispectrum}

We analyse both the two- and three-point statistics for different  models using first a single reference catalogue, and second a set of ten reference catalogues.

 We are using in this work consistently the same initial conditions at scales larger than the cell size resolution with \texttt{BAM} realisations as in the corresponding reference simulation. This already reduces to a great extent the effect of cosmic variance when making a comparison among them. Nonetheless, we have seen that the lowest few modes in the power spectrum are still affected by cosmic variance through halo bias. 

Using only one reference simulation enables us to investigate the accuracy in reproducing the statistics of that particular realisation. However, we extend the study to ten reference simulations, as this provides insights  on the robustness of the method against different realisations.

\begin{figure}
\hspace{0.5cm}
\includegraphics[width=7.5cm]{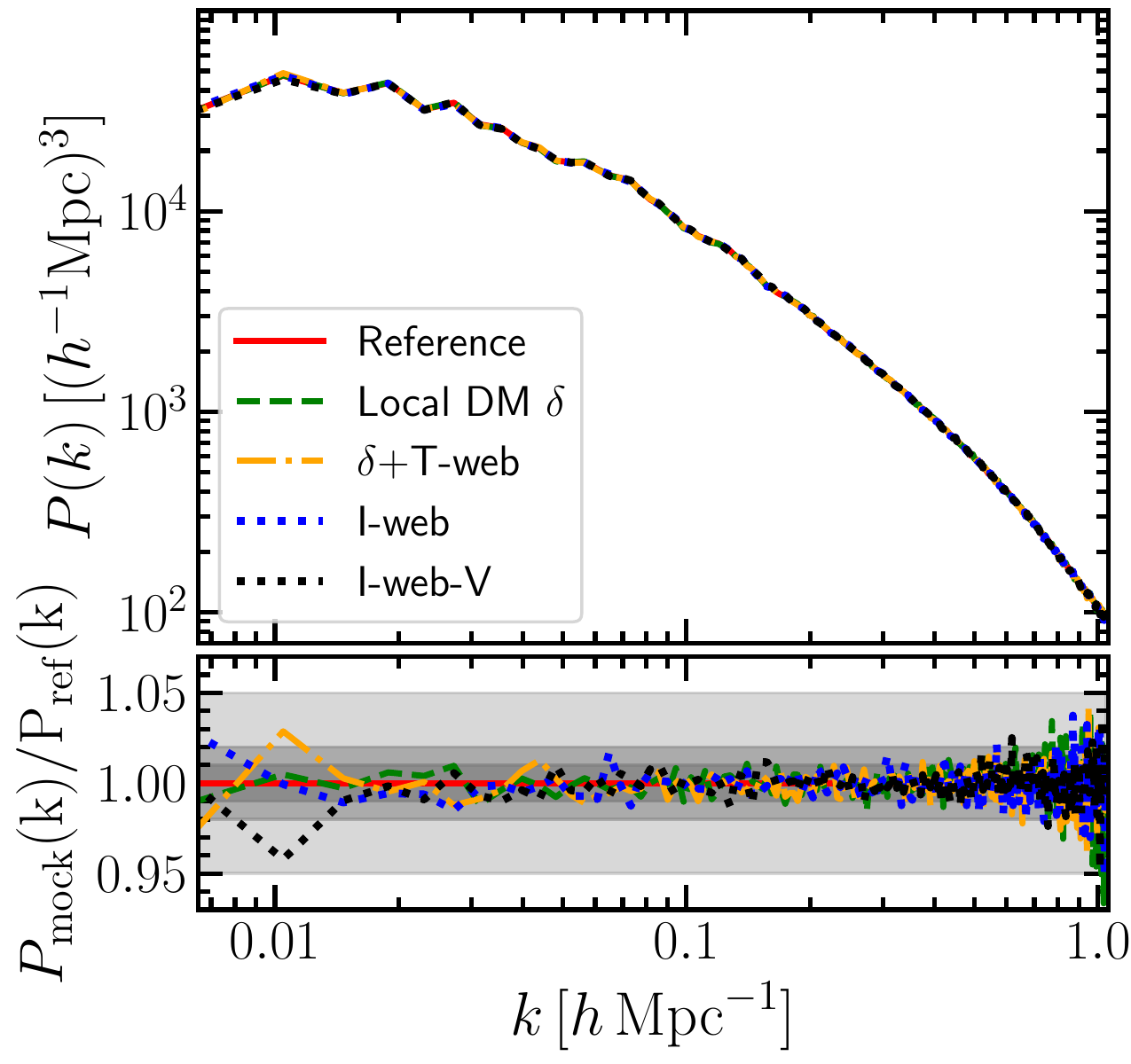}
\put(-95,180){One ref. Minerva sim.} 
\caption{Halo power spectra from  one reference Minerva simulation in red, and from \texttt{BAM} realisations with bias models given by $\delta$-, $\delta$+T-, and I-web using the same initial conditions (ICs). On the bottom panel corresponding  ratios w.r.t. the reference are shown. The different shaded areas stand for 1, 2, and 5\% error bars. See  \S\ref{sec:invariants} for PT-web and PT-web-q.}
\label{fig:PK}
\end{figure}

\begin{figure}
\hspace{0.5cm}
\includegraphics[width=7.5cm]{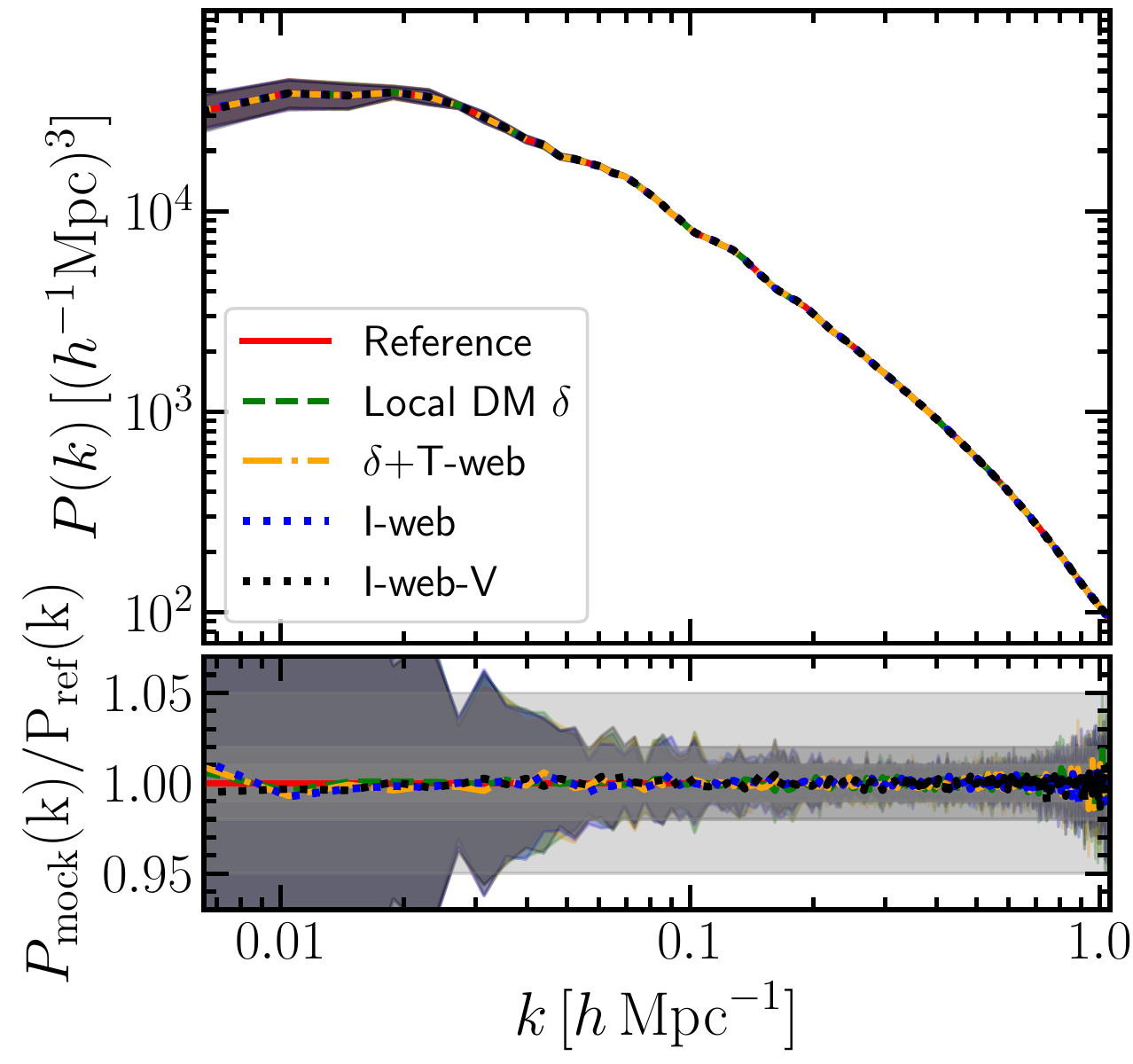}
\put(-95,180){Ten ref. Minerva sim.}
\caption{Same as Fig.~\ref{fig:PK} but based on ten reference Minerva simulations with different seed perturbations in the initial conditions showing the corresponding mean and 1-$\sigma$ contours for each case (they overlap significantly). }
\label{fig:PK10}
\end{figure}

The power spectra corresponding to different \texttt{BAM} runs w.r.t. one reference catalogue are shown in Fig.~\ref{fig:PK}. We have shown that we can reach with all the considered models an accuracy within 2\% in each mode, yielding residuals\footnote{Defined as the sum of the absolute difference between the reference  and the \texttt{BAM} catalogue's power spectra in the whole $k$-mode range up to 70\% of the Nyquist frequency.} below 1\%.
The results considering different realisations yield compatible power spectra for the various cosmic web classification methods (see Fig.~\ref{fig:PK10}).
 In  particular, we show results after the method is well converged with  the local $\delta$, the $\delta$+T-web,  the  I-web,  and the I-web-V model. The same level of accuracy is achieved for the PT-web, PT-web-q, and $\lambda$-web models in terms of the two-point statistics. We need to go to higher order statistics to find differences.

The three-point statistics is  commonly used to constrain the bias \citep[see e.g.][]{1997MNRAS.290..651M,2007PhRvD..76h3004S,2012PhRvD..86h3540B,2012MNRAS.420.3469P,2014PhRvD..90l3522S,2015MNRAS.450.1836K,2015JCAP...10..039A,2015JCAP...05..007B,2015JCAP...09..029A,2017MNRAS.465.1757G}.

\begin{figure*}
\begin{tabular}{c}
\hspace{-.9cm}
\includegraphics[width=1.07\textwidth]{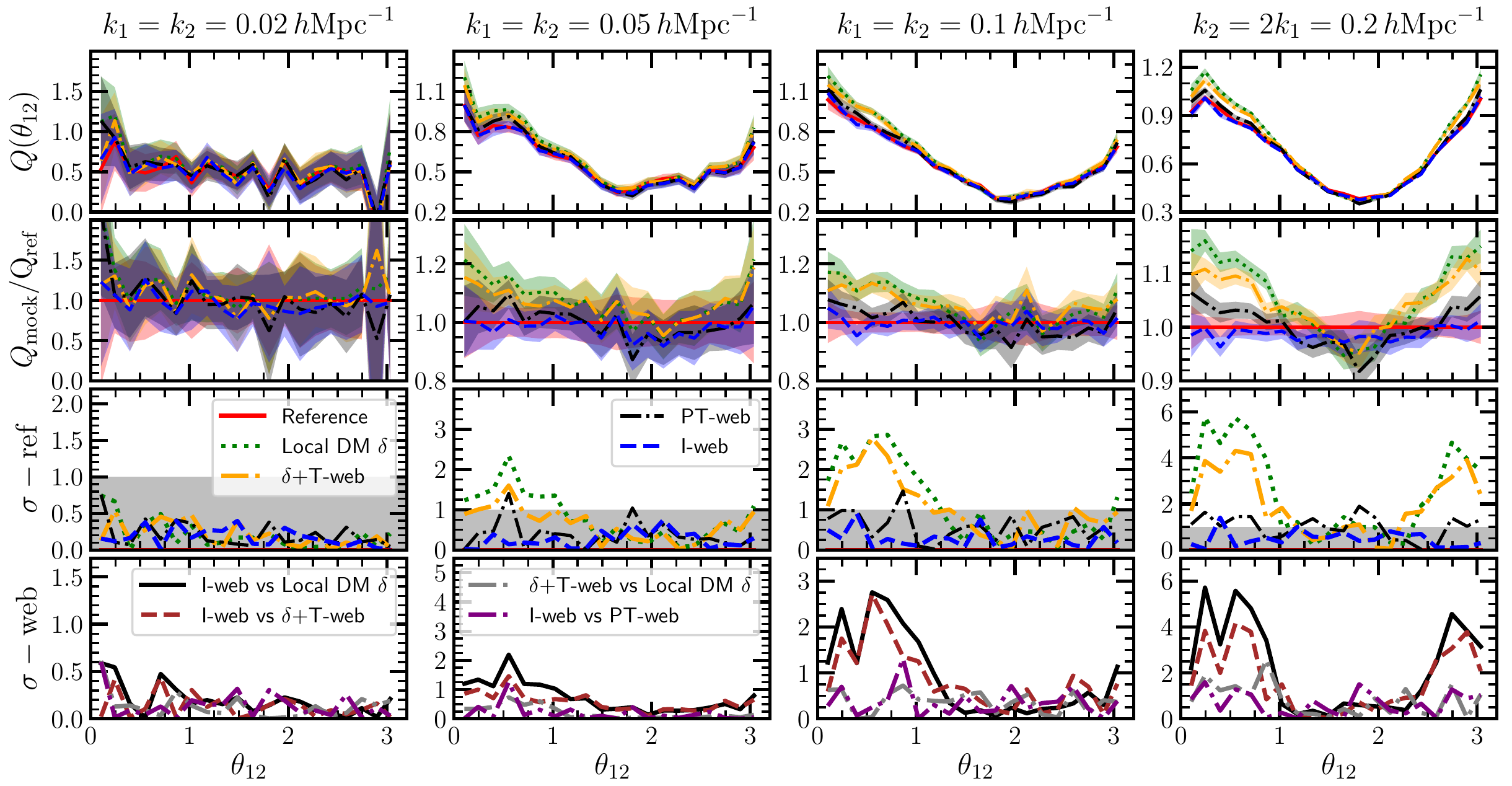}
\put(-300,290){One ref. Minerva sim.}
\end{tabular}
\caption{Reduced bi-spectra of the reference $N$-body halo catalogue (red solid) and the \texttt{BAM} mocks with the bias relation including: only $\delta$ (green dotted), $\delta$+T-web (dot-dashed orange), PT-web: $\{\delta,I_4,I_5\}$, and the I-web $\{\delta,I_2,I_3\}$. The corresponding ratios are shown in the second row. The third row shows the sigma difference: $\sigma_{\rm ref}(\text{X-web})=| Q_{\text{X-web}}-Q_{\rm ref}|/(\sqrt{2}\sigma)$, with $\sigma$ being the standard deviation extracted from the 300 Minerva catalogues. The last row shows the difference between two X and Y \texttt{BAM} mocks  $\sigma_{\rm web}(\text{X},\text{Y})=|\sigma_{\rm ref}(\text{X-web})-\sigma_{\rm ref}(\text{Y-web})|$, as indicated in the legend. The largest evidence for non-local bias is found for the I-web:  $\sigma_{\rm web}( \text{I-web},{\delta})$. This is a very conservative estimate, since the halo realisations share the same underlying dark matter field above a resolution of $3\,h^{-1}$ Mpc.} 
\label{fig:BK}
\end{figure*}

\begin{figure*}
\includegraphics[width=0.8\textwidth]{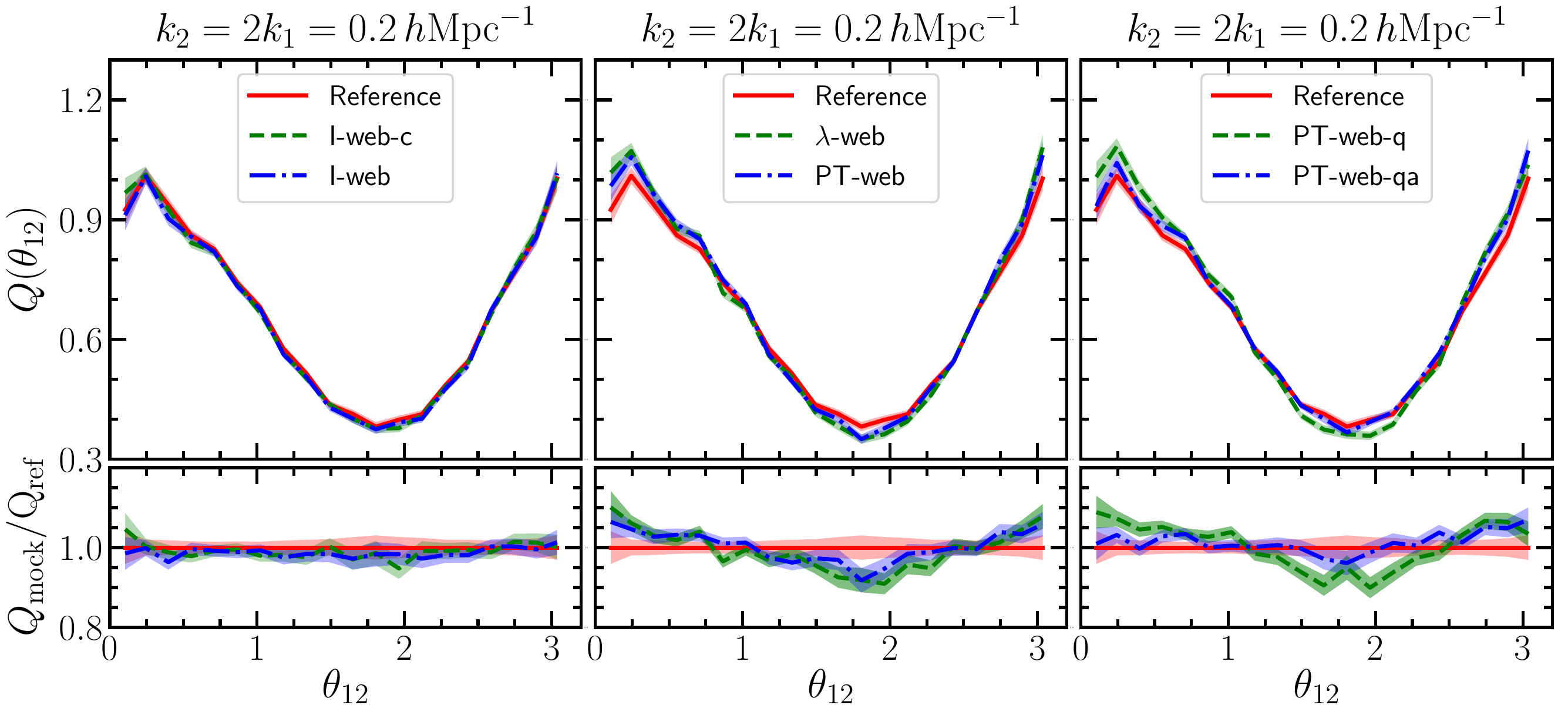}
%\put(-267,145){One ref. Minerva sim.}
\put(-240,185){One ref. Minerva sim.}
\caption{Reduced bi-spectra of one reference $N$-body halo catalogue (red solid) and the \texttt{BAM} mocks with the bias relation including on the left:  the I-web-c: $\{\delta,I_2,I_3,I_1I_2\}$  (green dashed) and the I-web: $\{\delta,I_2,I_3\}$  (blue dash-dotted); in the middle: $\lambda$-web: $\{\lambda_1,\lambda_2,\lambda_3\}$ and  PT-web: $\{\delta,I_4,I_5\}$; and on the right: PT-web-q: $\{\delta,I_4\}$ vs PT-web-qa: $\{\delta,I_4,s
^2(\alpha)\}$ (with 1-$\sigma$ contours from the 300 Minerva reference catalogues).} 
\label{fig:BK3}
\end{figure*}

\begin{figure*}
\rotatebox[]{0}{\includegraphics[width=0.8\textwidth]{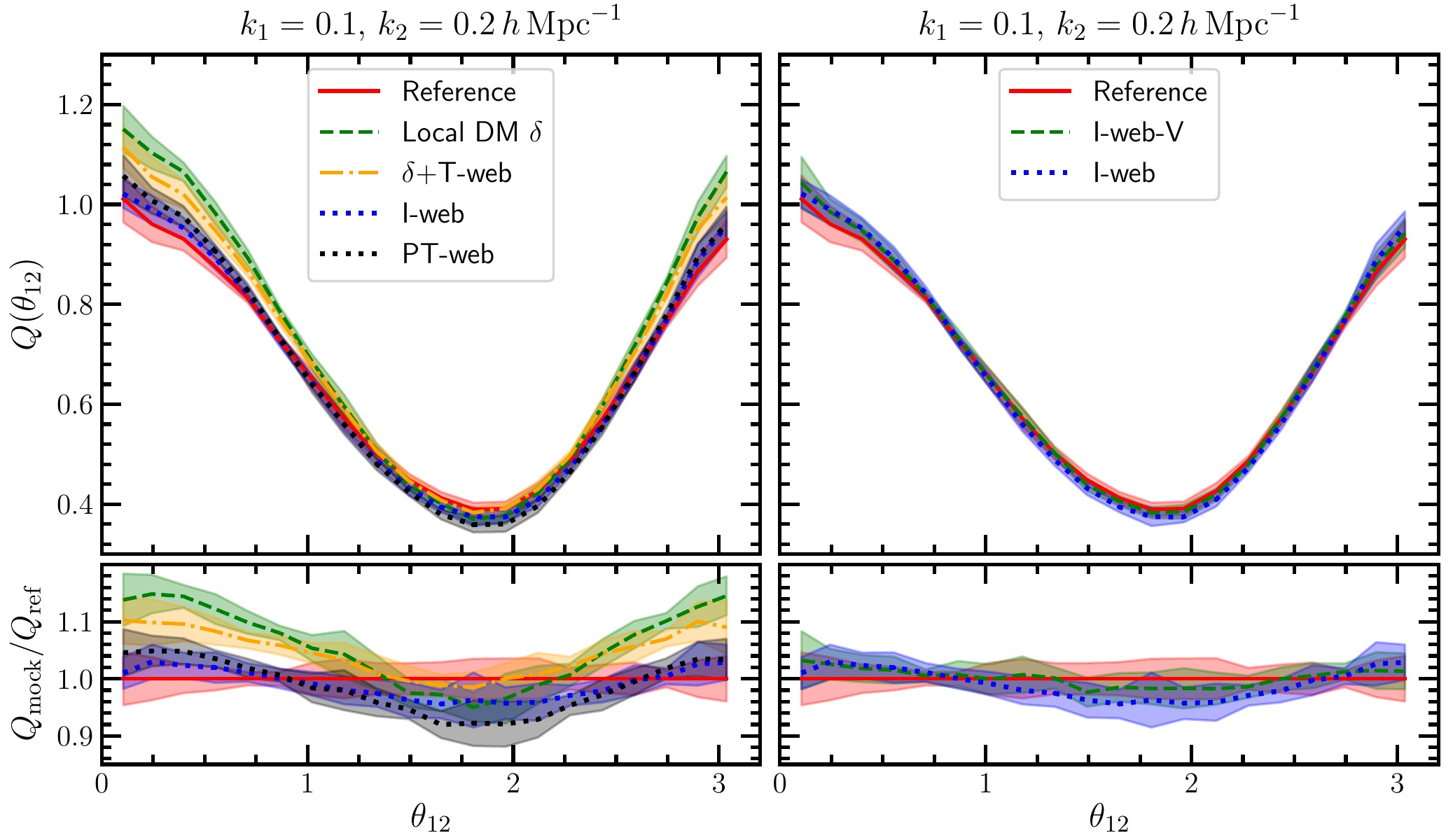}}
%\put(-145,210){Ten ref. Minerva sim.}
\put(-135,170){Ten ref. Minerva sim.}
\caption{Left panel: comparison between the bi-spectra corresponding to ten Minerva simulations and $\delta$, $\delta$+T-web, PT-web and I-web calculations (analogous to Fig.~\ref{fig:BK3}, including 1-$\sigma$ contours from ten realisations for each case). Right panel: the same as left panel showing I-web vs. I-web-V.}
\label{fig:BK10}
\end{figure*}

\begin{figure}
\hspace{0.05cm}
\includegraphics[width=8.5cm]{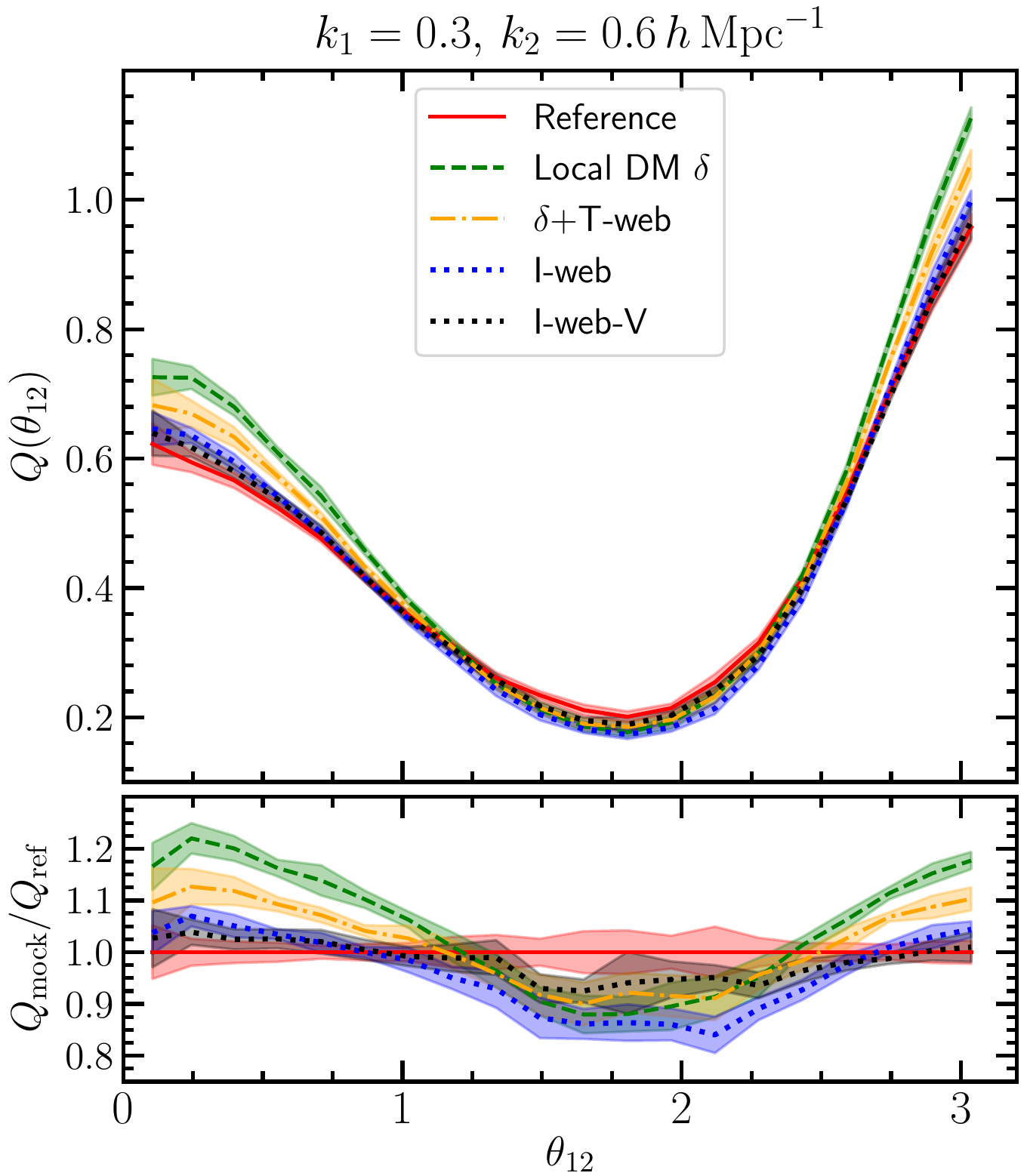}
%\put(-130,220){Ten ref. Minerva sim.}
\put(-150,180){Ten ref. Minerva sim.}
\caption{Same as Fig.~\ref{fig:BK10} at a smaller scale configuration.}
\label{fig:BK11}
\end{figure}

It is  investigated here in Fourier space through  the reduced bi-spectrum $Q(\theta_{12})$  as a function of the normalised angle $\theta_{12}$ between the $\vec k$-vectors $\vec k_{1}$ and $\vec k_{2}$ (see Figs.~\ref{fig:BK}- \ref{fig:BK11}).
 Comparing \texttt{BAM} runs with only local density information  vs I-web, we find  evidence for non-local bias at the 4.8 $\sigma$ confidence level (an information gain of  $\sim$ 3.4 $\sigma$ over $\delta$+T-web), being very conservative (see \S\ref{sec:stats}).  One should note that this confidence level is based on the assumption  of different seed perturbations, and hence it carries an overestimation of uncertainty due to cosmic  variance. Accordingly, the statistical significance of a non-local bias detection is expected to be stronger
  using more accurate covariance matrices with smaller uncertainties corresponding to the same initial conditions.
 
In particular, we consider triangle configurations in Fourier space constrained by the following two sides:  $|\vec k_{1}|=0.02,0.05,0.1\,h$ Mpc$^{-1}$ \& $|\vec k_{2}|=|\vec k_{1}|$; and $|\vec k_{1}|=0.1\,h$ Mpc$^{-1}$ \& $|\vec k_{2}|=2|\vec k_{1}|$, as a function of the $\theta_{12}$ angle.

 The bi-spectra at the largest scales are all compatible with each other within error bars  (configuration corresponding to $|\vec k_{1}|=0.02$ \& $|\vec k_{2}|=|\vec k_{1}|$,  on the left in Fig.~\ref{fig:BK}).
 It is remarkable,  that already at large scales with configurations of $|\vec k_{1}|=0.05\,h$ Mpc$^{-1}$ \& $|\vec k_{2}|=|\vec k_{1}|$ non-negligible differences can be found, as we  further analyze below.
 These differences increase towards smaller scales (see right panels).
 We stop at the configuration of $|\vec k_{1}|=0.1\,h$ Mpc$^{-1}$ \& $|\vec k_{2}|=2|\vec k_{1}|$, since this is the typically shown one, as being relevant to BAO and RSD analysis. Also higher $k$ configurations become computationally more expensive, and we plan to extend this work including velocity and short range non-local bias terms, for which smaller scales become interesting.

 Fig.~\ref{fig:BK3} shows the comparison between I-web vs  I-web-c, PT-web vs $\lambda$-web, and PT-web vs PT-web-qa.
 The left panel shows the similarity from considering or neglecting the cross term $I_1I_2$. As we argued before, $I_1I_2$ is well represented by $I_3$, which effectively   includes the third order non-local bias  terms $\delta\,s^2$ and $s^3$. From this we conclude that we can save in general the usage of an additional field. Although further analysis towards small scales will have to investigate this. The middle panel confirms that \texttt{BAM} is actually solving Eq.~(\ref{eq:qvdep}), as very different generating functions, which are supposed to approximately coincide in the final bias terms they produce, do actually have very similar power spectra and bi-spectra. The role of ellipticity and prolatness seems to be moderate in this statistics at those large scales.
 However, the difference between $I_4$ and $\alpha$ is complementary, adding more information on anisotropic clustering at second order. This is  demonstrated in the additional test shown in the right panel, which also partially explains the difference between the PT-web and the I-web models, and further analysed below. In the latter comparison the third order cross term $\delta\,s^2$, i.e., a further specification on the third order anisotropic clustering, can also play a role. It is tempting to expect \texttt{BAM} according to Eq.~(\ref{eq:qvdep}) to be able to construct $s^2(\alpha)$ from knowing $\delta$ and $I_4$, since $s
^2=I_4-\delta^2/3$. However, a greater accuracy is achieved when explicitly providing $s^2$ due to the low number of bins used. Adding a redundant variable in this case effectively increases the number of bins and provides more accurate results.   We  get bi-spectra, which are significantly closer to the reference (compare the --PT-web-q-- green to the --PT-web-qa-- blue dashed-dotted line on the right panel of  Fig.~\ref{fig:BK3}).

  We notice that $\lambda$-web does not fully characterise $s^3$ since $I_5$ needs an additional $I_3$ term which is not included in this model.

 We compute the  standard deviation w.r.t. the reference simulation as: $\sigma_{\rm ref}(\text{X-web})\equiv|Q_{\text{X-web}}-Q_{\rm ref}|/(\sqrt{2}\,\sigma)$, where we have calculated the covariance matrices of the bi-spectra from the reference simulation, and assume the same derived $\sigma$ error bars  apply for the \texttt{BAM}  realisations based on our previous experience \citep[][]{2020MNRAS.491.2565B}.
 Since the theoretical model in this case has a comparable error bar to the different models, we adopt a conservative procedure to estimate the significance of our measurements. In particular, we define the tidal field based non-local bias detection with the various  X-web models, as  $\sigma_{\mathcal T}(\text{X-web})\equiv \sigma(\text{X-web})-\sigma({\delta\text{-web}})$. The information gain of  the I-web w.r.t. the T-web is expressed via $\sigma(\text{I-web},\text{T-web})=\sigma(\text{I-web})-\sigma(\text{T-web})$. The significances $\sigma_{\mathcal T}(\text{X-web})$ and $\sigma(\text{I-web},\text{T-web})$ shown in  the lower panels of Fig.~\ref{fig:BK} are  positive, indicating that there is a detection of non-local bias and an information gain.
 By averaging over the angle range, which shows most differences ($0.25<\theta_{12}<0.75$), we find $\sigma_{\mathcal T}(\text{I-web})=4.8 \sigma$, $\sigma_{\mathcal T}(\text{T-web})=1.4 \sigma$, and hence $\sigma_{\rm web}(\text{I-web},\text{T-web})=3.4 \sigma$.
 
 This implies that we have a clear detection of the non-local bias in the bi-spectrum with the I-web, and a considerable information gain w.r.t the T-web.

\subsubsection{The cosmic web based on the velocity shear tensor}

\label{sec:shear}

In this section, we present calculations with \texttt{BAM} using the I-web based on the velocity shear instead of the tidal field tensor, i.e. the I-web-V model. The velocity field is taken from the Minerva reference simulation using a nearest-grid-point mass assignment scheme to define it on the same mesh as the density field. This model will effectively include not only terms in the first row of Eq.~(\ref{eq:bias}) but also terms in the second row for which $\theta\neq\delta$.

In particular, we consider the following variables for \texttt{BAM} based on the  I-web-V: \{$\delta$, $I_1(\mbi v)$, $I_2(\mbi v)$, $I_3(\mbi v)$\} with a binning of \{200, 100, 100, 70\}. This is comparable to the one of the I-web based on the tidal field tensor.

In fact, we find very similar bi-spectra which are not strongly affected by cosmic variance for the considered configurations (see left panel in Fig.~\ref{fig:BK10}).

A more careful analysis, comparing with velocity shear based realisations hints towards a slightly higher precision when relying on the I-web-V (see right panel in Fig.~\ref{fig:BK10}).

Since the results relying on the tidal field were already compatible with the reference simulation in the power spectrum and in the considered bi-spectrum configurations for the I-web case, we extend the three point statistics analysis to smaller scales.

Fig.~\ref{fig:BK11} shows that the additional information contained in the velocity field can provide a more precise bias description. In particular, the bi-spectra considering configurations of $k_2=2k_1 \sim0.6\hminv$  are, in general,  compatible within 1-$\sigma$ with the reference simulation when considering the velocity shear, as opposed to considering only the tidal field tensor.
Still, some slight discrepancy between the reference bi-spectrum and the I-web-V \texttt{BAM} realisations can be found. This could potentially be improved with an adequate binning or maybe relying on a larger number of reference simulations.
Further investigation on this is left for future work.

 \subsection{Statistical significance of the bias models}
 
 \label{sec:stats}

\begin{figure}
\hspace{-0.25cm}
\includegraphics[width=0.49\textwidth]{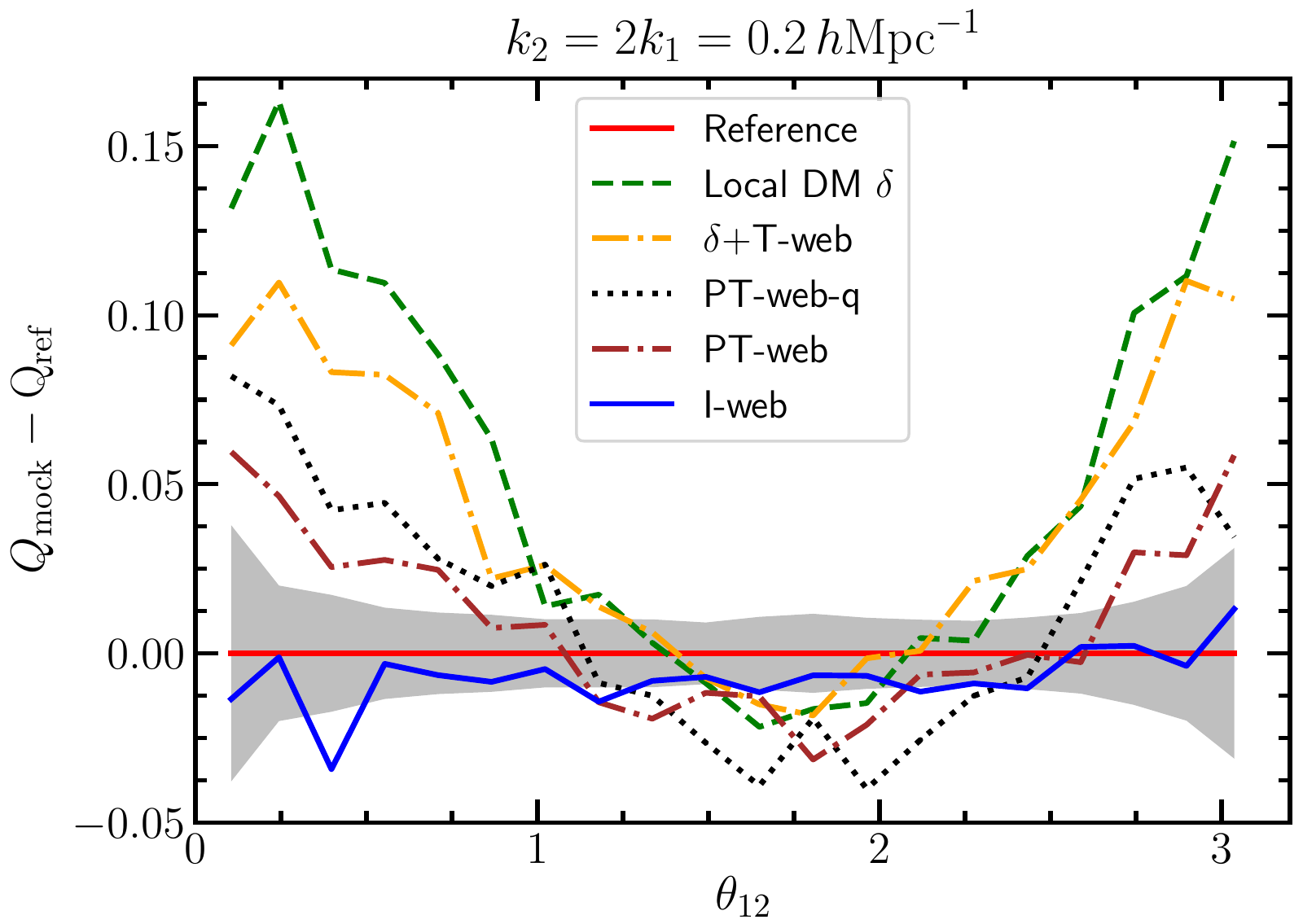}
\put(-145,83){One ref. Minerva sim.}
\caption{Difference of reduced bi-spectra of the reference $N$-body halo catalogue (red solid) and the \texttt{BAM} mocks with the bias relation including:  only $\delta$ (green dotted), $\delta$+T-web (dashed-dotted yellow), I-web: $\{\delta,I_2,I_3\}$, PT-web: $\{\delta,I_4,I_5\}$, and  PT-web-q: $\{\delta,I_4\}$. The grey area corresponds to the  standard deviation from computing the bi-spectra for 300 Minerva simulations.} 
\label{fig:BK2}
\end{figure}

\begin{table*}
\begin{tabular}{c|c|c|c|c|c|c|c|c|c|c|c|c|}
\cline{2-13}
& \multicolumn{12}{c|}{Bi-spectrum configuration}\\
\hline
\multicolumn{1}{|c|}{\multirow{2}{*}{Bias model}} & \multicolumn{3}{c|}{$k_1=k_2=0.02$} & 
\multicolumn{3}{c|}{$k_1=k_2=0.05$} & 
\multicolumn{3}{c|}{$k_1=k_2=0.1$}  & 
\multicolumn{3}{c|}{$k_1=0.1$ \& $k_2=0.2$}\\
\cline{2-13}
\multicolumn{1}{|c|}{} &$\chi^2$ & $P(X>\chi^2)$ & MR & $\chi^2$ & $P(X>\chi^2)$ & MR  & $\chi^2$ & $P(X>\chi^2)$ & MR  & $\chi^2$ & $P(X>\chi^2)$ & MR \\
\hline
 \multicolumn{1}{|c|}{Local DM $\delta$} &  0.17 &  99\% & \cellcolor{lightgray!30}{\bf N} & 2.0 &  0.4\% & \cellcolor{gray!60}{\bf Y} & 4.5 & $\sim$0\% & \cellcolor{gray!60}{\bf Y} & 20.2  & $\sim$0\% & \cellcolor{gray!60}{\bf Y}  \\
\hline
 \multicolumn{1}{|c|}{$\delta$+T-web}  & 0.13 &  99\% & \cellcolor{lightgray!30}{\bf N} & 1.14 & 29.4\% & \cellcolor{lightgray!30}{\bf N} & 3.37 & $\sim$0\% & \cellcolor{gray!60}{\bf Y} & 11.7 &  $\sim$0\% & \cellcolor{gray!60}{\bf Y}  \\
\hline
\multicolumn{1}{|c|}{PT-web-q}  & 0.15 & 99\% & \cellcolor{lightgray!30}{\bf N} & 0.61  & 90\%  & \cellcolor{lightgray!30}{\bf N} & 1.13 & 30\% & \cellcolor{lightgray!30}{\bf N} & 6.16 & $\sim$0\% & \cellcolor{gray!60}{\bf Y} \\
\hline
\multicolumn{1}{|c|}{PT-web} & 0.2 & 99\% & \cellcolor{lightgray!30}{\bf N} & 0.5 & 96\% & \cellcolor{lightgray!30}{\bf N} & 0.9  & 58\% & \cellcolor{lightgray!30}{\bf N} & 2.5 & $\sim$0\% & \cellcolor{gray!60}{\bf Y} \\
\hline
\multicolumn{1}{|c|}{I-web} & 0.1 & 99\% & \cellcolor{lightgray!30}{\bf N} & 0.2 & 99\% & \cellcolor{lightgray!30}{\bf N} & 0.3 & 99\% & \cellcolor{lightgray!30}{\bf N} & 0.7 & 85\%  & \cellcolor{lightgray!30}{\bf N} \\
\hline
\end{tabular}
\caption{\label{tab:tab1} Individual statistics for the different bias models on the corresponding bi-spectra: the $\chi^2$, the probability of a sampled $X$ being greater than the measured $\chi^2$ and the model rejection MR \{N: No, Y: Yes\}. We assume a model can be rejected when the measured $\chi^2$ is further than the 99\% probability of being within the $\chi^2_{\rm{dof}}$ distribution (dof is the number of degrees of freedom, number of bins in our case). The cell color gray indicates that  the model is rejected w.r.t. the reference, while light gray means that it cannot be rejected, with the I-web being the only model remaining green throughout all considered configurations.}
\end{table*}

\begin{table*}
\begin{tabular}{c|c|c|c|c|c|c|c|c|c|c|c|c|}
\cline{2-13}
& \multicolumn{12}{c|}{Bi-spectrum configuration}\\
\hline
\multicolumn{1}{|c|}{\multirow{2}{*}{Bias models}} & \multicolumn{3}{c|}{$k_1=k_2=0.02$} & 
\multicolumn{3}{c|}{$k_1=k_2=0.05$} & 
\multicolumn{3}{c|}{$k_1=k_2=0.1$}  & 
\multicolumn{3}{c|}{$k_1=0.1$ \& $k_2=0.2$}\\
\cline{2-13}
\multicolumn{1}{|c|}{} & $\Delta\chi^2$ & BF & E & $\Delta\chi^2$ & BF & E & $\Delta\chi^2$ & BF & E & $\Delta\chi^2$ & BF & E \\
\hline
 \multicolumn{1}{|c|}{I-web vs local DM $\delta$} & 0.08 & 1.04 & \cellcolor{lightgray!30}{\bf N} & 1.81 & 2.47 & \cellcolor{lightgray!30}{\bf N} & 4.21 & 8.21 & \cellcolor{gray!60}{\bf Y}1 & 19.5 & 17402 & \cellcolor{gray!60}{\bf Y}4 \\
\hline
 \multicolumn{1}{|c|}{I-web vs $\delta$+T-web}   & 0.04  & 1.02 & \cellcolor{lightgray!30}{\bf N} & 0.94  & 1.60 & \cellcolor{lightgray!30}{\bf N} & 3.1 & 4.72 & \cellcolor{gray!60}{\bf Y}1 & 11.06 & 252 & \cellcolor{gray!60}{\bf Y}4 \\
\hline
 \multicolumn{1}{|c|}{I-web vs PT-web-q} & 0.06 & 1.03 & \cellcolor{lightgray!30}{\bf N} & 0.41 & 1.23 & \cellcolor{lightgray!30}{\bf N} & 0.87 & 1.54 & \cellcolor{lightgray!30}{\bf N} & 5.48 & 15.47 & \cellcolor{gray!60}{\bf Y}3  \\
\hline
 \multicolumn{1}{|c|}{I-web vs PT-web} & 0.06 & 1.03 & \cellcolor{lightgray!30}{\bf N} & 0.30 & 1.16 & \cellcolor{lightgray!30}{\bf N} & 0.64 & 1.37 & \cellcolor{lightgray!30}{\bf N} & 1.81 & 2.50 & \cellcolor{gray!60}{\bf Y}1 \\
\hline
\end{tabular}
\caption{\label{tab:tab2}  Comparison of bias models based on the corresponding bi-spectra, showing the differences $\Delta\chi^2$, the corresponding Bayes factors BF, and the evidence grades of the difference in the models: \{N: No evidence, Y1: Substantial, Y2: Strong, Y3: Very Strong, Y4: Decisive\}. The evidence grades are taken from \citet[][]{Jeffreys:1939}. The cell color light  gray means that there is no evidence that a model is preferred over the I-web model, while gray stands for evidence against the various models  w.r.t. the I-web model. All cases become red, meaning that all models are disfavoured w.r.t. the I-web.}
\end{table*}

To compute  the statistical significance of the different models we rely on the $\chi^2$ statistics  (see Fig.~\ref{fig:BK2}). 

 We assume that the bi-spectrum error bars are given by the set of 300 Minerva simulations. We should note, that the real error bars must be somewhat lower, since the cosmic variance is suppressed to some extent by using the same initial conditions at a down-sampled resolution.

In Tab.~\ref{tab:tab1} we show individual measurements of $\chi^2/$dof (number of degrees of freedom, the amount of $\theta_{12}$ bins in our case) and its interpretation from the frequentist point of view. We show what is the probability of finding a sample of the $\chi^2$ distribution $X$ greater than the one given by the mocks $P(X>\chi^2)$. We assume that the  computed $\chi^2$ is not drawn from the same distribution, if its probability is not within the $99\%$ of the distribution. We found that the only mock satisfying this condition was the one produced in the I-web case.  

In Tab.~\ref{tab:tab2} we show the comparison between $\chi^2/$dof, and the interpretation in terms of Bayes factors. Since we assume that both the priors and the probability of the data (also called the evidence) are the same in all of the cases, the Bayes factor reduces to the likelihood ratio. The grades of evidence are taken from \citet{Jeffreys:1939}. We find that for the largest scales there is no evidence for preferring any model over another. However, towards smaller scales the I-web parametrisation evidence increases, as compared to any other model.
The I-web shows a clear preference w.r.t. the local density and the $\delta$+T-web models already for the configuration $k_1=k_2=0.1\,h$Mpc$^{-1}$.
The PT-web and PT-web-q models compete with the I-web until the $k_1=0.1$ \& $k_2=0.2\,h$Mpc$^{-1}$ configuration is achieved, for which we find evidence that the I-web is required to fit the $N$-body reference halo catalogue.

 From this analysis, we infer that the I-web runs are the only ones, which are indistinguishable  from the reference (at the considered scales). We also find that the PT-web model does not match the reference simulation, but performs better than the PT-web-q model, which contains only up to quadratic terms. This study implies that we have a considerable gain from including second order non-local bias ($I_4$) terms (black dotted vs green dashed lines, and an additional more moderate information gain from including third order non-local bias ($I_5$) terms (see brown dashed-dotted vs black dotted lines  in Fig.
~\ref{fig:BK}). A considerable information gain is obtained from using the I-web, which has a more accurate description of the anisotropic clustering as discussed above  (see blue solid vs brown dashed-dotted lines in Fig.
~\ref{fig:BK}).

\section{Discussion and Conclusions}

\noindent 

In this work we have investigated the characterisation of the gravitational deformation tensor through its invariants  and its impact on the distribution of dark matter tracers.   We have shown how to link the latter with the usual characterisation of the cosmic web relying on combinations of the eigenvalues and their connection to the perturbative halo bias expansion  \citep[see, e.g.,][]{1986ApJ...304...15B,2002MNRAS.329...61S,2009JCAP...08..020M,2012PhRvD..85h3509C,2012MNRAS.420.3469P,2012PhRvD..86h3540B,2014MNRAS.440..555P,2014PhRvD..90l3522S,2017MNRAS.472.3959M,2018JCAP...07..029A,2018JCAP...09..008L,2021PhRvD.103l3550E}.
We have done this  along the lines of \citet[][]{2019PhRvD.100d3514S},  populating the dark matter field at the Eulerian field level.
This has enabled us to reproduce the results of full gravity calculations based on $N$-body simulations to unprecedented accuracy in the two- and three-point statistics of dark matter tracers. To this end we have relied on the  Bias Assignment Method \citep[\texttt{BAM}][]{2019MNRAS.483L..58B,2020MNRAS.491.2565B,2020MNRAS.493..586P}, which acts as a  physically motivated supervised machine learning method able to combine generating  functions to produce halo counts on a regular mesh. This approach has been followed by similar concepts relying on deep learning. While they are very promising \citep[see e.g.,][]{2019arXiv190205965Z} when trying to reproduce the summary statistics on large cosmic volumes, they  do still not reach the precision of the \texttt{BAM} algorithm  \citep[][]{2019PhRvD.100d3515K}.
Machine learning methods are not only more difficult to interpret in terms of physical insights, but also very sensitive to the definition of the cost function they aim at minimising, and other set-up parameters like the number of layers, etc. Hereby, the non-local dependencies are difficult to track. Physical models help to increase the accuracy, as demonstrated by combining the Zel'dovich approximation with machine learning \citep[][]{2019PNAS..11613825H}. The \texttt{BAM} approach aims in fact at providing the maximum physical information to minimize  the uncertainties encoded in a single kernel (as a function of $k$ in Fourier space), which is extracted within a Markov Chain Monte Carlo rejection algorithm, learning from the reference simulation. Hence, deep learning approaches could benefit from the insights provided in this work.

\noindent  
 The bi-spectrum analysis shows that already at configurations of $k_1=k_2=0.05\hminv$  the invariants become necessary.  
 This accuracy increases when using the invariants of the tidal shear tensor towards smaller scales as expected from having a more complete anisotropic bias description.
 
 A precise description of anisotropic clustering has turned out to be crucial on large scales confirming its importance found in previous work \citep[][]{2019MNRAS.489.2977R}.  
 
In summary, we have succeeded to find a consistent picture at scales relevant to BAO and RSD analysis. In forthcoming publications, the connection to the galaxy distribution, redshift space distortions, and covariance matrices going down to lower halo masses will be presented (Balaguera-Antol\'inez, Kitaura et al in prep.).
From recent studies showing that an accurate fit to the two-  and three-point statistics implies accurate covariance matrices \citep[][]{2018MNRAS.480.2535B}, we expect that the corresponding four-point statistics will also be well reproduced with \texttt{BAM} relying on the I-web.
In fact, previous versions of the \texttt{BAM} code using the T-web already reproduced well the covariance matrices from $N$-body simulations using the information of the power spectra up to $k=0.2\hminv$  \citep[][]{2020MNRAS.491.2565B}.

\noindent 
These findings suggest that one might find very complete galaxy bias descriptions  based on a few terms constructed with  invariants of the tidal field or velocity shear tensor. In particular, one could easily extend the bias model suggested by  \citep[][]{1993ApJ...413..447F,1993ApJ...417..415C,2012MNRAS.427...61A}   with something like: 
$\rho_{\rm g}\propto{h(\{d_i\})\,f\left(\sum_i\,a_i\,(c_i+ g(\eta_i))^{b_i}\right)}$,
where $f$, $g$, and $h$ are some appropriate functions, $a_i$, $b_i$, $c_i$, and $\{d_i\}$ are a set of bias factors, and $\eta_i$ being,   e.g.,  \{$\delta(I_1),I_2,I_3,(I_1I_2)$\}). The function $h$ typically models the suppression  of the appearance of galaxies towards low densities through a threshold or  decaying exponential \citep{1984ApJ...284L...9K,2014MNRAS.439L..21K,2014MNRAS.441..646N}.
Some interesting  cases can be found for $f$ being the identity function and $g$ being an exponential, or reversing the roles of $f$ and $g$.
We leave a study of the velocity  shear, vorticity, as  well as density short range terms as a function of redshift for future work.
This work  also suggests that galaxy evolution and formation studies, which are recently relying on cosmic  web classifications, could potentially benefit from a different angle. Parameter regions defined by the multidimensional space spanned by the invariants 
could become the sights  to look at, to  identify common properties of galaxies.
We have worked out, how   one could associate those parameter regions to different physical  properties.
As an example, one could relate a particular I-web parameter region to a combination of cosmic web types with their associated probabilities.
As a further application, the I-web is expected to improve the halo mass reconstruction, previously relying on the T-web  \citep[][]{2015MNRAS.451.4266Z}.

\noindent 
We conclude  that the invariants of the gravitationally evolved tidal field tensor at Eulerian coordinates are able to characterize the cosmic web, and the halo bias terms down to the Eulerian field level to great accuracy. Thus,  they could be useful to extract cosmological information from the next generation of galaxy surveys.

\section*{Data availability}

The data underlying this article will be shared on reasonable request to the corresponding author.  

\section*{Acknowledgments}
The authors thank A.~S\'anchez, P.~McDonald, R.~Sheth, R.~Angulo, R.~van de Weygaert, C.~Zhao,  C.-H.~Chuang, C.~Dalla Vecchia, and J.~A.~Rubi\~no-Mart\'in for  discussions. %Special thanks to A.~S\'anchez for providing the Minerva simulations.  
Special thanks to AS and CDV for kindly providing the Minerva simulations, and CZ and CHC  for the bi-spectrum  code. FSK and ABA acknowledge the IAC facilities and  the Spanish Ministry of Economy and Competitiveness (MINECO) under the Severo Ochoa program SEV-2015-0548, AYA2017-89891-P and  CEX2019-1124000920-S grants. FSK also thanks the  RYC2015-18693 grant. FS thanks the University of Padova and the IAC.
MPI thanks the AYA2012-39702-C02-01 and ERC-SG 716151 (BACCO) grants.

\bibliographystyle{mnras}
\bibliography{lit.bib}

\appendix

\section{Relation between the invariants and the cosmic web classification}
\label{sec:cwinv}

In this appendix we study the relation between the cosmic web classification and the invariants of the tidal field tensor, defined in Eq.~(\ref{eq:princi}).

In particular, we consider different cases based on the I-web restricted to the information provided by $I_1$, $I_2$, and $I_3$  \citep[we also consider without loss of generality a threshold eigenvalue of zero, as in][]{2007MNRAS.375..489H}:
\begin{itemize}
    \item $I_3>0$: $\lambda_1\lambda_2\lambda_3>0$, which leaves two options, either knots: \{$\lambda_i>0$ $\forall i$\}, or sheets:   \{$\lambda_1>0,\lambda_i<0$  for $i=2,3$\} 
    \begin{itemize}
        \item $I_2<0$: $\lambda_1\lambda_2+\lambda_1\lambda_3+\lambda_2\lambda_3<0$. Since $\lambda_1>0$ for both knots and sheets, we can multiply with $\lambda_1$, hence obtaining: $\lambda_1\lambda_2\lambda_3+\lambda_1^2\lambda_2+\lambda_1^2\lambda_3<0$, but since $I_3>0$, it follows that  $\lambda_3<-\lambda_2$. This is  only  accomplished for sheets.
        \item $I_2>0$ is equivalent to $I_1>\lambda_1-\lambda_2\lambda_3/\lambda_1$. On the other hand  $I_1>\lambda_1$ for knots, but $I_1<\lambda_1$ for sheets, since for such cosmic web types  $\lambda_2+\lambda_3<0$.
    \end{itemize}
    \item $I_3<0$: $\lambda_1\lambda_2\lambda_3<0$, i.e., either voids: \{$\lambda_i<0$ $\forall i$\}, or filaments:   \{$\lambda_i>0$, for $i=1,2$, $\lambda_3<0$ \}  
    \begin{itemize}
        \item $I_2<0$: since for both cases $\lambda_1\lambda_2>0$, we have $\lambda_1\lambda_3+\lambda_2\lambda_3<0$. Also, for both $\lambda_3<0$. Thus, $\lambda_1>-\lambda_2$, which is only accomplished for filaments.
        \item $I_2>0$: Since for both $\lambda_3<0$, we get $I_1<\lambda_3-\lambda_2\lambda_3/\lambda_1$. On the other hand, for voids $\lambda_1+\lambda_2<0$, hence, $I_1<\lambda_3$. And for filaments $\lambda_1+\lambda_2>0$ yielding $I_1>\lambda_3$.
    \end{itemize}
\end{itemize}
In summary, the connection between the T-web and the I-web in terms of the invariants of the tidal field tensor is given by: 
\begin{itemize}
    \item knots: $I_3>0$ \& $I_2>0$ \& $I_1>\lambda_1$
    \item filaments: $I_3<0$ \& $I_2<0$ $||$ $I_3<0$ \& $I_2>0$  \& $\lambda_3<I_1<\lambda_3-\lambda_2\lambda_3/\lambda_1$
    \item sheets: $I_3>0$ \& $I_2<0$ $||$ $I_3>0$ \& $I_2>0$  \& $\lambda_1-\lambda_2\lambda_3/\lambda_1<I_1<\lambda_1$
    \item voids: $I_3<0$ \& $I_2>0$ \& $I_1<\lambda_1$
\end{itemize}
The introduction of a threshold other than zero will shift these relations. 
One should  note, that the classification into different cosmic web types apparently requires the specification of certain eigenvalues, but if the density $I_1$ is known over its whole range of values, the combination with the sign    of the other two invariants ($I_2$ and $I_3$) fully constrains the different cases. In any case, from the characteristic polynomial each eigenvalue can be computed with the knowledge over the first three invariants. 
What has been found here also applies to the velocity shear classification (V-web)  by substituting the tidal field tensor with the shear tensor.  The difference is that the V-web based on the velocity  field beyond linear theory, effectively includes additional terms which  we classified into the $F_{\rm shear}(\vec v(\vec r))$ and $F_{\rm curl}(\vec v(\vec r))$ terms in Eq.~(\ref{eq:bias}), as we investigate in \S\ref{sec:shear}.

It is interesting to note, non-theless, that the T-web (and V-web) has been a useful tool, as it carries  more information
  than the density alone, involving the  second order non-local bias $I_4$ (through $\delta$ and $I_2$), and partial information on  the third order $s^3$ term.
  However, from these calculations we find that the T-web (and V-web) constrains only a sub-region of the parameter space spanned by the I-web, involving only $I_1,I_2,I_3$ for very restricted cases. Thus the T-web cannot account for anisotropic dependencies such as ellipticity, prolatness, or the anisotropic parameter. It can neither account for the non-local bias terms $\delta s^2$, nor properly for $s^3$, since $I_5$ is not constrained by the T-web classification.

%%%%%%%%%%%%%%%%%%%%%%%%%%%%%%%%%%%%%%%%%%%%%%%%%%

%%%%%%%%%%%%%%%%% APPENDICES %%%%%%%%%%%%%%%%%%%%%

%%%%%%%%%%%%%%%%%%%%%%%%%%%%%%%%%%%%%%%%%%%%%%%%%%

% Don't change these lines
\bsp	% typesetting comment
%{\begin{tabular}[c]{@{}c@{}}$\mathcal{T}_4$\\ $i=1$ \end{tabular}} & 3 &    340& 2994 & 10 &  70$|$90  & XX & XX & XX & XX& 1.48  & 1.15$|$X    \\ \hline

\label{lastpage}

\end{document}